\documentclass[aps,prb,twocolumn,showpacs]{revtex4}
\usepackage{amsmath,amsthm}  
\usepackage{epsf}
\usepackage{amssymb}  
\usepackage{graphicx}

\usepackage{bbm}
\usepackage{bm}
\newcommand{\ket}[1]{| #1 \rangle}
\newcommand{\kket}[1]{| #1 \rangle\rangle}
\newcommand{\bra}[1]{\langle#1  |}
\newcommand{\bbra}[1]{\langle\langle \tilde{#1}  |}
\newcommand{\rb}[1]{\left( #1 \right)}

\newcommand{\ew}[1]{\langle #1 \rangle}
\newcommand{\be}{\begin{eqnarray}}
\newcommand{\ee}{\end{eqnarray}}

\newcommand{\eq}[1]{Eq.~(\ref{#1})}
\newcommand{\fig}[1]{Fig.~(\ref{#1})}
\newcommand{\vecu}{\mathbf{u}}
\newcommand{\vecv}{\mathbf{v}}

\newcommand{\vecW}{\mathbf{W}}

\begin{document}
\title{Feedback  Control of Waiting Times}
\author{Tobias Brandes$^1$ and Clive Emary$^2$}
\affiliation{  $^1$ Institut f\"ur Theoretische Physik,   Hardenbergstr. 36,   TU Berlin,   D-10623 Berlin,   Germany.\\ 
$^2$ Joint Quantum Centre Durham-Newcastle, School of Mathematics and Statistics, Newcastle University, Newcastle upon Tyne, NE1 7RU, United Kingdom.
}
\date{\today{ }}
\begin{abstract}
Feedback loops are known as a versatile tool for controlling transport in small systems, which usually have large intrinsic fluctuations. Here we investigate the  control of a temporal correlation function, the waiting time distribution, under active and passive feedback conditions. We develop a general formalism and then specify to the simple unidirectional transport model, where we compare
costs of open loop and feedback control and use methods from  optimal control theory to optimize waiting time distributions. 
\end{abstract}

\maketitle

\section{Introduction}
Feedback is a closed loop control scheme where some part of the dynamics of a system is recycled in order to achieve a certain control goal \cite{Schoell2016,WM2009,Jacobs2014,Zhangetal2015,FBreview2013}. 
Often, this goal consists in stabilizing the overall system dynamics, reducing fluctuations, or preparing certain states \cite{Wiseman94,Horoshko_1997,Korotkov99,Korotkov01,Ruskov02,Korotkov05,Combes06,Jacobs07,Combes08,Combes_2010,Ueda2010,Blum10,Gillett10,Ralph11,Vijay12,Riste12,Inoue13,Wallraff_2014}. 
This is  particularly challenging in small systems with large intrinsic fluctuations, where one has to combine methods from
non-equilibrium statistical mechanics, measurement and (quantum) information theory to properly design and analyse the control action.  

An important issue is to assess under which conditions feedback is useful and more efficient and effective than other control schemes such as open loop control or an {\em a priori} optimization of system parameters. At present,  there is no complete framework within thermodynamics or some kind of resource theory \cite{resourcetheory} that would fully resolve this issue, and the best way forward seems to be the study of well-defined physical setups where feedback operations are expected to be beneficial. 

In this paper, we concentrate on minimal models for transport, motivated by recent experimental realizations \cite{FBexperiment16} of feedback loops  in mesoscopic solid state systems, i.e., quantum dots. 
An interesting finding there was the reduction of the overall fluctuations of the electric current due to the feedback loop at large times \cite{Bra10}, in the form of the freezing of the full counting statistics $p(n,t)$, the probability of $n$ charges being transfered across the dot during a time interval $[0,t]$. In contrast,  short-time correlations as quantified by the waiting time distribution $w(\tau)$ were found to be essentially unaltered by the feedback scheme. 

We have confirmed these observation in our calculations, and taken this as a motivation to turn the question around by asking, more generally, for feedback schemes (`protocols' for the time-dependent steering of system parameters as introduced below) that have a strong impact on $w(\tau)$. In simple words, the feedback goal then is to reduce short--time fluctuations, i.e., the stochasticity in the random time intervals $\tau$ between two (quantum) jumps. 

Our choice of waiting times as the subject of studying feedback control of correlation functions has several reasons. First, in Markovian processes they are a  natural choice, in particular in 
simple situations where the system is reset to one and the same state after each jump  (as described by re-newal theory \cite{renewaltheory}). Often enough, there is a close connection of $w(\tau)$ to other correlations functions such as the $g^{(2)}(\tau)$ function in quantum optics \cite{Emary2012}. 
  In physical chemistry \cite{CaoSilbey} and applied mathematics (queueing theory), a vast literature on various aspects of waiting times exists. Second, waiting times form the basis of quantum trajectories that have been developed for the master equations in quantum optics since the 1980s \cite{Carmichael_02}. The unraveling of such master equations, i.e., the splitting into jump and non-jump parts, then automatically leads to the path-integral like formal series solution that has turned out to be most suitable for a phenomenological introduction of measurement-based feedback, even with delay \cite{Schoell2016}. Finally, in solid state physics waiting times have recently emerged as a powerful tool to analyse transport \cite{Bra08,EspositoLindenberg08,Welack08}, also beyond the simple Markovian limit \cite{Albert11,Albert12,Thomas13,Rajabi13,Dasenbrook14,Thomas14,Haack14,Albert14,Sothmann14,Dasenbrook15,Albert16,Dasenbrook16}. This last aspect is particularly promising as it might offer a way to introduce feedback  control, at least as passive control \cite{EG2014}, in highly non-Markovian situations.

The distinction between measurement-based (active) and passive  feedback,  cf. Fig. (1),   is particularly important in the quantum regime. There,  passive feedback (sometimes called coherent then) avoids issues related to the quantum measurement problem by building the feedback loop as part of the total system. Coherent feedback has been introduced very successfully in quantum optics \cite{Lloyd2000,Zhang2010,Nori2013,Caretal2013,HSCK2014,GPS2014,Schulze2014,Kabuss2015,Hein2015,Grimsmo2015,Kabuss2016}  and in coherent quantum transport recently \cite{EG2014}. 

In general, a key question in all feedback schemes is to determine the efficiency of the control loop, also in comparison with open loop control. Much progress has been achieved over the past few years in the analysis of feedback from a thermodynamic perspective, based on concepts such as entropies, mutual information, and  modifications of fluctuation relations or the various formulations of the second law \cite{SU2008,SU2010,HP2011,SU2012,AS2012,MR2012,ES2012,HSP2013,SSEB2013,Tas2013,HE2014,HBS2014,BS2014,SS2015,SSBJ2014,Hor2015,HJ2015}. Applied to concrete control scenarios, this analysis however often requires certain assumptions, e.g., a bipartite splitting into system and controller \cite{HE2014}, or the maintainance of a certain system state as the control goal \cite{HJ2015}. 

One feature of the waiting times feedback scheme introduced below is the possibility to perform feedback control without modifying the exchange fluctuation relation. Costs and efficiencies for feedback conditioned upon the previous quantum jump  can also  be optimized by using  methods from  optimal control theory, which in our view is a successful path towards a  phenomenological understanding of this kind of feedback. Still, from  a microscopic perspective the most transparent way to interpret an active feedback scheme is a mapping onto an equivalent passive realisation of that scheme within a physical setting that can then be analysed. This clearly lacks the generality with that statements can then be made, since such a mapping is not unique, but it often has the advantage of being more realistic in terms of a concrete physical implementation and experimental realization \cite{SSEB2013,Pekola15}.

The outline and the main results of this paper are as follows. We first introduce the formal framework and various feedback protocols in Sect. II and  III.  In Sec. IV we introduce the waiting times, before analysing a particular control scheme, feedback conditioned on the previous jump, and ability to influence waiting times and full counting statistics in Sect. V. It turns out that within this particular feedback scheme, many of the calculations  are very similar to the Markovian case except for time-dependencies which always occur, in contrast to open loop control, as differences $t-t_n$ between the present time $t$ and the time $t_n$ of the previous jump. 
In Sect. 
\ref{optimization} we show that this property can be used to optimize the feedback protocol with methods from classical optimization theory. The costs functionals appearing 
in the optimization also turn out to be a very efficient tool for assessing and comparing the `costs' of various feedback schemes. We use these in order to carry out a detailed comparison between open and closed loop feedback in Sect.  \ref{section_comparison}, before carrying out a thermodynamic analysis based on fluctuation relations, the  Kullback--Leibler divergence, and 
finally a  passive feedback analogon for an analysis in terms of  Shannon entropies in Sect. \ref{section_thermo}.

\section{Method}
Our starting point is an open  physical system -- classical or quantum -- interacting with several reservoirs, measurement and feedback devices. The state of the system at time $t$ is given by a 
reduced system density operator $\rho(t)$. In the classical case, this is a vector of probabilities in the space of system states (assumed as discrete here), in the quantum case one has additional coherences. 
We decompose $\rho(t)$ according to 
\be\label{rhon}
\rho(t) &\equiv&  \sum_{n=0}^{\infty}\rho^{(n)}(t),
\ee
where $n$ is the total number $n$ of (quantum) jumps in $[0,t]$ defined via stochastic trajectories, i.e. sequences  $ \{ x_n \} = x_n,...,x_1$ of 
transitions among the system states. These transitions are of  type $l_i$ and occur at time $t_i$, and we use the shorthand  notation $x_i=(l_i,t_i)$. 

Next, we introduce conditioned and unnormalized  density operators $\rho(t |\{ x_n \})$ in the sense that the $\rho^{(n)}(t)$ are given in terms of a `path integral';
\be\label{unravel}
\rho^{(n)}(t) = 
 \sum_{l_1=1,...,l_n=1}^M   \int_0^{t} dt_n ...\int_0^{t_{2}}dt_{1} 
\rho(t |\{ x_n \}).
\ee  

For a  Markovian, time-independent  quantum master equation without any form of control, the $\rho(t |\{ x_n \})$ can be expressed explicitly in the usual unraveling procedure:  
The reduced density operator obeys 
\be
\label{eofmotion}
\dot{\rho}(t) = \mathcal{L}\rho(t),\quad \mathcal{L}=\mathcal{L}_0+\mathcal{L}_1,\quad \mathcal{L}_1=\sum_{k=1}^M \mathcal{J}_k,
\ee
where the total Liouvillian $\mathcal{L}$ is split into two  superoperators:  $\mathcal{L}_1$ describes $M$ different types of jump processes, whereas $\mathcal{L}_0$ is the generator for the time evolution 
 $S_t\equiv  e^{\mathcal{L}_0 t}$ between the jumps. In this case, \eq{unravel} is the usual unraveling with 
\be
\label{rho_cond}
\rho(t |\{ x_n \}) \equiv S_{{t}-t_n} \mathcal{J}_{l_n}  S_{t_n-t_{n-1}} \mathcal{J}_{l_{n-1}}  ... \mathcal{J}_{l_1}  S_{t_1} \rho_{\rm in}
\ee
and $\rho^{(0)}(t) = S_t  \rho_{\rm in}$, which technically follows  from the solution of \eq{eofmotion} in the interaction picture with respect to  $\mathcal{L}_0$. Here and in the following, 
$ \rho_{\rm in}\equiv \rho(t=0)$ denotes the initial density operator at time $t=0$.

\subsection{Feedback model}
We now introduce control operations in the following way:  the system parameters at time $t>0$ are continuously modulated depending on the values $x_i$ of the previous jump events. Starting the time evolution at $t=0$, the system is monitored until the first jump occurs at time  $t=t_1$. 
During that period, the  Liouvillian  becomes time-dependent, 
\be
\mathcal{L}(t) =\mathcal{L}^0(t) + \sum_{l=1}^M \mathcal{J}_{l}(t),\quad 0\le t \le t_1, 
\ee
and thus until the occurance of the first jump, the time evolution becomes
\be
\rho(t) &=& S(t)  \rho_{\rm in},\quad  S(t) \equiv T e^{\int_0^t dt' \mathcal{L}^0(t')},
\ee
where $T$ is the time ordering operator. In the course of the time evolution, all jump events $x_n$ are recorded and taken as parameters in the subsequent Liouvillians, which  read
($t_n\le t \le t_{n+1}$)
\be
\mathcal{L}(t|\{ x_n \}) &=&\mathcal{L}^0(t|\{ x_n \}) + \sum_{l=1}^M \mathcal{J}_{l}(t|\{ x_n \}).
\ee
The time evolution between the $n$th and $n+1$th jump is now generated by  
\be\label{Sdef}
S(t|\{ x_n \}) \equiv T e^{\int_{t_n}^t dt' \mathcal{L}^0(t'|\{ x_n \})},
\ee
as the solution of $ \frac{d}{dt}S(t|\{ x_n \}) =  \mathcal{L}^0(t|\{ x_n \})S(t|\{ x_n \}) $  with $S(t_n|\{ x_n \})=\hat{1}$ (unity operator). 

For  example, with $n=2$ jumps, the density  operator conditioned on $x_1,x_2$  reads 
\be\label{2example}
\rho(t |\{ x_2 \}) &=& S(t| \{ x_2 \})  \mathcal{J}_{l_2}(t_2|\{ x_1 \})   S(t_2| \{ x_1 \}) \nonumber\\
&\times &  \mathcal{J}_{l_1}(t_1)  S(t_1)  \rho_{\rm in},
\ee
and the corresponding $n$-resolved density operator, i.e., the $n=2$ term in \eq{unravel},
is obtained by summation/integration over the jump variables $x_1,x_2$, 
\be\label{2exampleb}
\rho^{(2)}(t) = \sum_{l_1=1,l_2=1}^M   \int_0^{t} dt_2 \int_0^{t_{2}}dt_{1} \rho(t |\{ x_2 \}).
\ee
Iterating \eq{2example},  we obtain the sequence 
\be\label{rhocondgeneral}
\!\!\!\!
\rho(t |\{ x_n \}) &=& S(t| \{ x_n \})  \mathcal{J}_{l_n}(t_n|\{ x_{n-1} \})   S(t_n| \{ x_{n-1} \})... \nonumber\\
&\times &        \mathcal{J}_{l_2}(t_2|\{ x_1 \})   S(t_2| \{ x_1 \})      \mathcal{J}_{l_1}(t_1)  S(t_1)  \rho_{\rm in},
\ee
with  (non)-jump time-evolutions for the unraveling of $\rho(t)$ according to \eq{unravel} in which each step in conditioned upon the previous ones. 

The form \eq{rhocondgeneral} is a key result in the formulation of active, measurement based control as introduced here. This formulation is phenomenological in the sense that, in contrast to the expression \eq{rho_cond} without control, it has not been derived from a total, microscopic Hamiltonian for the system including all reservoirs, measurement, and control devices. 
As in  \eq{rho_cond}, the $\rho(t |\{ x_n \})$ are given by a sequence of jump and non-jump time-evolutions. Here, this sequence is generated by superoperators which
themselves in general depend on the trajectories $\{ x_n \}$ and thus are random quantities.  As we show below, this leads to a number of powerful control schemes, some of which have been successfully applied in the past already.

\section{Feedback protocols}\label{sectionfp}
The specific form of the jump operators $ \mathcal{J}_{l}(t|\{ x_{n} \})$ defines a particular feedback procedure, called `protocol' in the following. In general, this opens numerous ways to control a stochastic process, possibly also including modulation of the parameters in the Hamiltonian in the quantum case.

\subsection{Open loop control \label{sectionfpopenloop}}
The simplest case is open loop control with jump operators
\be
  \mathcal{J}_{l}(t|\{ x_{n} \})
  =
  \mathcal{J}_{l}(t)
\ee
that are just modulated as a function of time and thus do not depend on the stochastic process itself. 
Needless to say that this already leads to a vast variety of control schemes, cf. section \ref{section_comparison} below. For example, a periodic modulation of parameters leads to Floquet-type (quantum) master equations. More general time-dependencies could be optimized with methods of optimal control theory (see below) in order to reach a specific control target, such as adiabatic control with slow pulses.

\subsection{Time-versus-number feedback \label{sectionfptvn}}

The next example is a dependence  
\be
  \mathcal{J}_{l}(t|\{ x_{n} \})
  =
  \mathcal{J}_{l}(t,n)
\ee
on time $t$ and the number $n$ of quantum jumps only. A protocol of this nature has been proposed and realized experimentally recently \cite{Bra10,Bra15,FBexperiment16}. More will be said about this scheme after \eq{FBexperiment} and in section \ref{section_unidirectional}.

\subsection{Feedback conditioned on the previous jump}\label{sectionfpconditioned}
Next, we introduce an efficient feedback protocol that in the main focus for the rest of this paper. This is feedback  conditioned on the previous jump event $x_n$ only, instead on the whole trajectory $x_n,x_{n-1},...,x_1$, i.e., 
$\mathcal{J}_{l}(t|\{ x_n \}) =\mathcal{J}_{ll_n}(t-t_n)$. Here and in the following, we already assumed that the protocol immediately starts (without delay) after the time $t_n$ of the previous jump, and used the jump-type $l_n$ as an additional index at the jump operators. 
The  conditioned density operators in \eq{rhocondgeneral} now become
\be\label{previousJ}
\rho(t |\{ x_n \}) &=& 
S_{l_n}(t-t_n)     \mathcal{J}_{l_n,l_{n-1}}(t_n-t_{n-1}) \times \\
& ...& 
 \mathcal{J}_{l_2,l_1}(t_2-t_1)  
S_{l_1}(t_2-t_1) \mathcal{J}_{l_1}(t_1)  S(t_1) \rho_{\rm in}.\nonumber
\ee
A feature of this protocol is that the same form of time--dependence in the control parameters is repeated over and over again after each jump.

One simple example of this type of protocol is a discontinuous change in the jump operators from $\mathcal{J}_l$ to $\mathcal{J}_l'$ after a fixed delay time:
\be\label{delayedcontrol}
  \!\!\!
  \mathcal{J}_{ll_n}(t-t_n) 
   &=& 
   \mathcal{J}_l \theta(\tau - t + t_n)
   +
   \mathcal{J}'_l\theta(t - t_n -\tau),
\ee
with $\theta(t)$  the unit step function, and where we have assumed a uniform delay time $\tau$ for all jump processes.  Jump operators of this form were employed in Ref.~[\onlinecite{Emary2013}], (see also Ref.~[\onlinecite{Wiseman94}]) to model delayed-feedback control in quantum transport. In  this context, the jump operators $\mathcal{J}_l$ were the ones of the original system, and $\mathcal{J}'_l= e^{\mathcal{K}_l}\mathcal{J}_l$ were the controlled jump operators where the original operator is followed by a control operation.  In the limit $\tau \to 0$, the first term in \eq{delayedcontrol} vanishes, and we are left with 
$
  \mathcal{J}_{ll_n}(t-t_n) 
  =
   \mathcal{J}'_l = e^{\mathcal{K}_l}\mathcal{J}_l
$, which is the instantaneous-control form of Wiseman and Milburn \cite{Wiseman94,WM2009}. In this way, we see that the Wiseman-Milburn feedback scheme maps onto an effective time-independent open-loop control problem, in which we simply design the control operations $e^{\mathcal{K}_l}$ to produce the desired modification of system behaviour.  A generalization of this scheme where the control operations $e^{\mathcal{K}_l}$ are chosen randomly 
has recently been proposed \cite{Daryanoosh2015}. 

We note that in practical terms, feedback conditioned on the previous jump is able to efficiently influence the waiting time distribution. In contrast, time-versus-number feedback was used to optimize the full counting statistics in the experiment \cite{FBexperiment16} but turned out to leave the waiting time distribution essentially unchanged, at least for small feedback strengths.

\section{Waiting times}

In the following, we will consider the feedback-conditioned-on-the-previous-jump protocol, and further restrict ourselves to jump superoperators  $\mathcal{J}_{l}$ that can be expressed as dyadic products multiplied by the rates $\gamma_l$ at which the jumps occur,
\be\label{Jketbra}
\mathcal{J}_{l}(t|\{ x_n \}) \equiv \gamma_l(t|\{ x_n \}) \kket{l}\bbra{l}.
\ee
This is a convenient notation where kets like $ \kket{l}$ denote  column  vectors  with dimension $d$  which is given by the number of real elements (including coherences in the quantum case) of the density operator $\rho$. Superoperators like $\mathcal{J}_{l}$ then act as $d\times d$ matrices and are conveniently represented in dyadic form. Note that  bras $\bbra{l}$ ($d$-dimensional row vectors) and kets  $ \kket{l}$  in \eq{Jketbra} in general need not to be orthogonal to each other. By convention, one  sets $\bbra{0}=(1,1,....,1,0,0...0)$ which represents the trace operation via the normalisation $ \bbra{0}{l}\rangle\rangle=1$, and one  defines $ \kket{0}$ as the representative of  $\rho_{\rm in}$ (the state at $t=0$).  
The form \eq{Jketbra} means that the character of the jump processes remains invariant under feedback, it is just the rates of the processes that are altered.

\subsection{Definitions}
Using this notation, we can re-write the conditioned density operator of \eq{rhocondgeneral} as
\be\label{waitcond}
\rho(t |\{ x_n \}) &=& S(t| \{ x_n \}) \kket{l_n} w_{l_nl_{n-1}}(t_n|\{ x_{n-1} \}) \times \nonumber\\
&\times & ...  w_{l_2l_{1}}(t_2|\{ x_{1} \}) v_{l_1}(t_1).
\ee
This defines the (feedback-conditioned)  waiting time distributions between a jump of type $l'$ at time $t_n$ followed by a jump of type $l$ at time $t>t_n$ 
\be\label{waitdef}
 w_{ll'}(t|\{ x_{n} \}) \equiv  \gamma_l(t|\{ x_n \}) \bbra{l} S(t|\{ x_n \}) \kket{l'}
\ee
via the matrix elements of the non-jump time evolution operators $S$.  Additionally,  
\be\label{vldef}
v_{l_1}(t_1) \equiv  \gamma_{l_1}(t_1) \bbra{l_1} S(t_1) \kket{0}
\ee
is the waiting time distribution for the first jump at time $t_1$ after initialization at $t=0$. 

The  waiting time distributions must be normalized to one when summed over all final jump types $l$ and integrated over all times $t\ge t_n$. Indeed, using the normalization $\bbra{0} l\rangle\rangle = \bbra{0} l'\rangle\rangle=1$ and $ S(t_n|\{ x_n \})= \hat{1}$,  one finds
\be
& &\sum_{l=1}^M \int_{t_n}^\infty dt  w_{ll'}(t|\{ x_{n} \}) \nonumber\\
&=& \int_{t_n}^\infty dt  \bbra{0} \left[ \mathcal{L}(t|\{ x_n \}) - \mathcal{L}^0(t|\{ x_n \})\right]  S(t|\{ x_n \}) \kket{l'}\nonumber\\
&=& -\int_{t_n}^\infty dt  \bbra{0} \frac{d}{dt} S(t|\{ x_n \})   \kket{l'} = \bbra{0} l'\rangle\rangle =1,
\ee
where in the total Liouvillian we used the vanishing of all column sums, $\bbra{0} \mathcal{L}(t|\{ x_n \})=0$, which expresses conservation of probability. 

\begin{figure}[t]
\centerline{\includegraphics[width=\columnwidth]{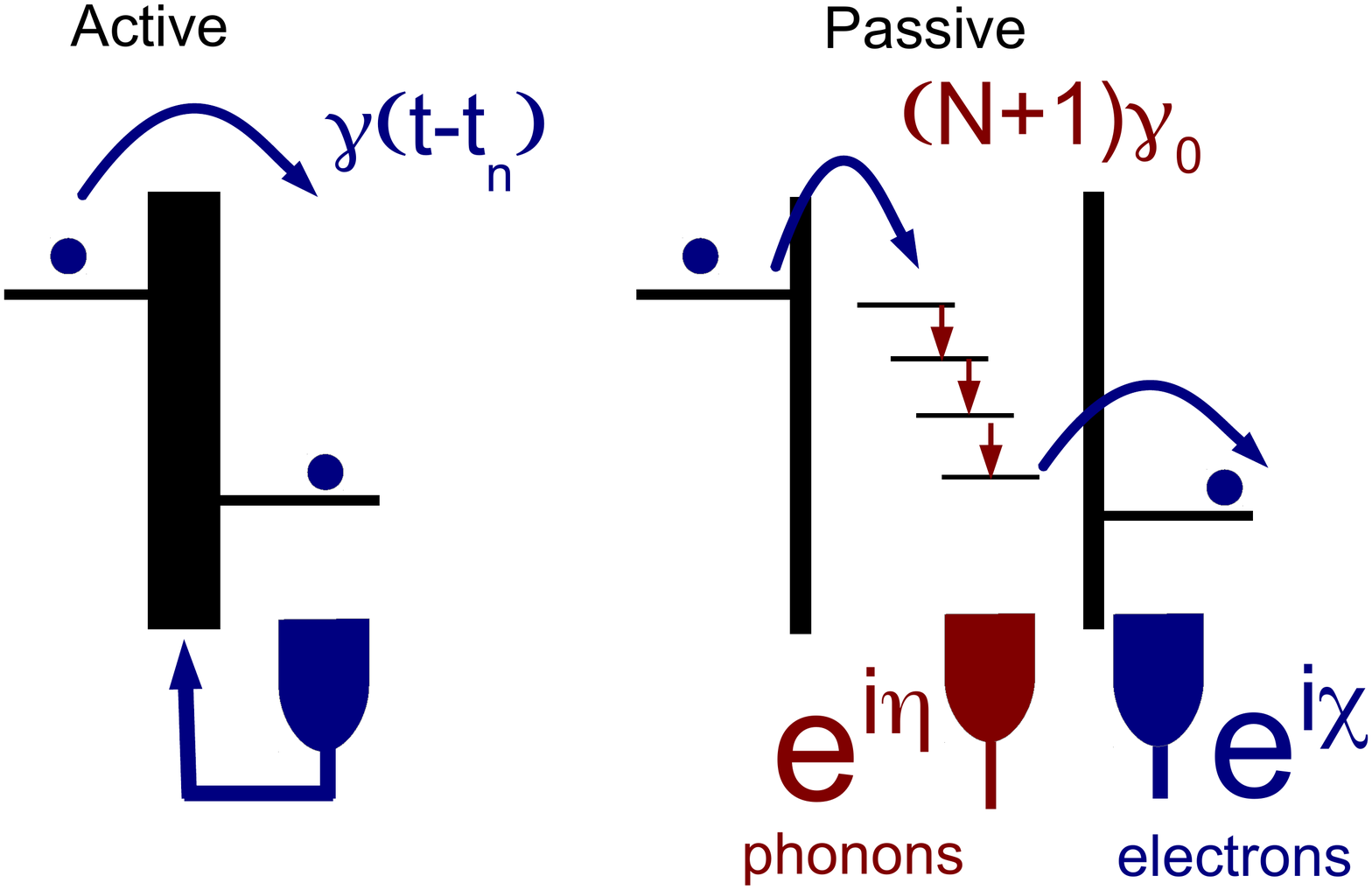}}
\caption[]{\label{waitingtime_fbscheme} LEFT: Scheme of active (measurement based) feedback control for the example of a unidirectional stochastic process, \eq{pntunidirectional}. The feedback protocol is realized via the single time-dependent rate $\gamma(t-t_n)$, starting again and again at the time $t_n$ of the previous jump, cf. \eq{previous}. RIGHT: Passive feedback scheme (`hardwiring') with series of $N+1$ inelastic  transitions (with phonon emissions from  quantum dot with $N$ levels) at identical rates $(N+1)\gamma_0$, and electrons leaving from the lowest dot level only,  to simulate an equivalent waiting time distribution without feedback from measurement devices, cf. Sect. \ref{section_passive}.}
\end{figure} 

\subsection{Examples}\label{section_examples}
At this stage, it is instructive to give a few instructive examples of the general expressions \eq{waitcond} and \eq{waitdef} before we proceed to specific feedback protocols. 

{\em Unidirectional stochastic process.---} This is defined by transitions between states $0\to 1\to 2 \to ...$ at (conditioned) time-dependent rates $\gamma(t| \{ x_{n} \})$ between $n$ and $n+1$ with $n\in \mathbb{N}$, where by convention the process starts with $n=0$ at time $t=0$. Consequently, the set of jump events $\{ x_{n} \}$ is uniquely defined by the times $t_n,t_{n-1},...,t_1$ only. Clearly, there is no additional structure apart from the probabilities $p(n,t)$ of the system being in state $n$ at time $t$.

Starting at $n=0$, the first jump operators are 
\be
\mathcal{J}_{1}(t_1)  &\equiv& \gamma(t_1) \ket{1}\bra{0},\quad
\mathcal{J}_{2}(t_2| t_1) \equiv \gamma(t_2|t_1) \ket{2}\bra{1}\nonumber\\
\mathcal{J}_{3}(t_3| t_2t_1) &\equiv& \gamma(t_3|t_2t_1) \ket{3}\bra{2},... 
\ee
where the kets (denoted as $\ket{n}$ here) simply form a Cartesian basis with dual basis  $\bra{n}$. Formally, the $S$--operators \eq{Sdef} are diagonal as 
$S(t|\{ t_n \})  = \exp  \left[-\int_{t_n}^t dt' \sum_{m=0}^\infty \ket{m} \bra{m} \gamma(t'|\{ t_n \}) \right]$. 
In \eq{unravel} we have the normalization $\bbra{0} \rho(t) =  \sum_{n=0}^{\infty}  p(n,t)=1$ with  $p(n,t)\equiv \bbra{0}\rho^{(n)}(t)$, and explicitly 
\be\label{pntunidirectional}
& &p(n,t) = \int_0^{t} dt_n ...\int_0^{t_{2}}dt_{1} 
e^{-\int_{t_n}^t dt'\gamma( t'|\{t_n\}) } \times\nonumber \\
&\times  & 
 w(t_n|\{ t_{n-1} \}) \times ...\times  w(t_2|\{ t_{1} \})  w(t_1),
\ee  
with the definition of the waiting time distributions ($n\ge 2$) 
\be\label{waituni}
w(t_n|\{ t_{n-1} \}) \equiv \gamma( t_n|\{t_{n-1}\}) e^{-\int_{t_{n-1}}^{t_n} dt'\gamma(  t'|\{t_{n-1}\}) }
\ee
and $ w(t_1)\equiv \gamma(t_1) e^{-\int_{0}^{t_1} dt'\gamma(t') }$.

In quantum transport, \eq{pntunidirectional} is a model for the counting statistics of a detector that counts transitions across a highly biased single tunnel barrier (without any additional internal structure like quantum dots energy level).
In this context, a time-versus-number protocol has been proposed and recently  realized experimentally \cite{Bra10,Bra15,FBexperiment16}.  Here the single rate  $\gamma(t|\{ x_n \})$ is taken to depend only on time $t$ and the number $n$ of quantum jumps.  In  Refs.~[\onlinecite{Bra10,Bra15,FBexperiment16}], the form taken was
\be\label{FBexperiment}
 \gamma(t,n) \equiv \gamma[ 1 + g (\gamma t - n)] 
 ,
\ee
where $\gamma$ is a target rate and $g>0$ a feedback parameter that is used to continuously adapt the jump rate depending on the current status of the system $n$, compared against a target status $\gamma t$.

Explicitly, the  waiting times
\be\label{wndef}
w_n(t_n,t_{n-1}) \equiv \gamma(t_n,n-1) e^{-\int_{t_{n-1}}^{t_n} dt' \gamma(t',n-1)}
\ee
now depend on two times (first and subsequent second jump) and  are conditioned upon the total number of jumps $n$ in the time interval between $[0,t_n]$ . 
We define a stationary waiting time distribution $w(\tau)$ as the $w_n$, \eq{wndef} weighted with the probabilities $p_n(t)$ in the limit $t\to \infty$,
\be\label{wtnscheme}
w(\tau) \equiv \lim_{t\to \infty} \sum_{n=0}^\infty p_n(t) w_n(t_n=t+\tau, t_{n-1}=t). 
\ee
The $w(\tau)$ are normalized, $\int_0^\infty d\tau w(\tau)=1$, which follows from $w_n(t+\tau,t)= -\frac{d}{d\tau}  e^{-\int_{t}^{t+\tau} dt' \gamma(t',n-1)} $ and the normalization $\sum_{n=0}^\infty p_n(t) =1$.
The explicit evaluation of  \eq{wtnscheme}  in Appendix A yields
\be\label{wtFBexp}
w(\tau) = \gamma e^{-\gamma \tau}\left( 1 + g \left( 1- \frac{\gamma\tau}{2} -\frac{(\gamma\tau)^2}{4}      \right)\right) + O(g^2),
\ee 
which for small feedback parameters $g\ll 1$ is very close to the (Poissianian) non-feedback distribution.
This is in agreement with the experiment \cite{Wagner_Haug_private2016}, where at the same time a strong reduction of the shot noise was found in the form of a feedback--frozen second cumulant \cite{FBexperiment16}, cf. \eq{C2freezing} in Sect. \ref{section_unidirectional}.

{\em Bidirectional stochastic process.---} This is defined by forward and backward rates $\gamma_\pm(t| \{ x_{n} \})$ between states $m$ and $m\pm 1$ with $m\in \mathbb{Z}$. The jump operators \eq{Jketbra} are defined
as 
\be\label{bidirectional}
J_{m\pm} (t|\{ x_n \}) \equiv \gamma_{m\pm}(t|\{ x_n \}) \ket{m\pm1}\bra{m}.
\ee
This is a minimal model for a discussion of detailed balance and fluctuation relations and will be analysed in more detail in Sect. \ref{section_bidirectional}.

{\em Single atom resonance fluorescence.---} 
This is an example of a two-level system with quantum coherences in the density operator $\rho$, which in absence of feedback control obeys a master equation
\be\label{Rabi}
\dot{\rho} = \gamma 
 (2 \sigma_- \rho \sigma_+ - \rho \sigma_+\sigma_- - \sigma_+\sigma_-\rho)  + i\frac{\Omega}{2}[\sigma_++\sigma_-,\rho],
\ee
where $\Omega$ is the Rabi frequency, $\gamma$ the spontaneous emission rate, and $\sigma_-\equiv \ket{-}\bra{+}$, $\sigma_+\equiv \ket{+}\bra{-}$. Writing  $\rho=(\rho_{++},\rho_{--},\rho_{+-},\rho_{-+})^T$ in vector form, the feedback conditioned jump operators becomes $J (t|\{ t_n \}) \equiv \gamma(t|\{ t_n \}) \kket{1} \bbra{1}$ with $\kket{1} = (0,1,0,0)^T$, $\bbra{1} = (1,0,0,0)$, and the  rate  $\gamma$  conditioned on the previous jump times $\{ t_{n} \}$. We note that  these times  could also be also used for conditioning the Rabi frequency $\Omega=\Omega (t|\{ t_n \})    $ (`Hamiltonian feedback') which enters the non-jump time-evolution operators $S$, \eq{Sdef}.

\section{Full counting statistics and waiting time feedback}
We now discuss the relation between waiting times and the full counting statistics under the previous-jump-conditioned feedback protocol of  Sec.~\ref{sectionfpconditioned}.
In this case, the trajectories can be written 
\be\label{previous}
\rho(t |\{ x_n \})
&=&  S_{l_n}(t-t_n) \kket{l_n} w_{l_nl_{n-1}}(t_n-t_{n-1} ) \times \nonumber\\
&\times & ...  w_{l_2l_{1}}(t_2-t_{1}) v_{l_1}(t_1),
\ee
with  $w_{ll'}(t-t') \equiv  \gamma_l(t-t') \bbra{l} S_{l'}(t-t') \kket{l'}$ replacing \eq{waitdef} for the waiting time distributions. 
Here, the non-jump time-evolution operators, \eq{Sdef}, are $S(t|\{ x_n \}) \equiv S_{l_n}(t-t_n)\equiv T \exp[\int_{t_n}^t dt' \mathcal{L}^0_{l_n}(t'-t_n)]   $ and thus only functions of the time differences
$t-t_n$, which we recognize by substituting $t'-t_n \to t'$ in  the integrand $\mathcal{L}^0_{l_n}(t'-t_n)$. 

An advantage of this protocol is that the same form of time--dependence in the control parameters is repeated over and over again after each jump. Technically, this has the advantage that we can immediately simplify \eq{previous} by Laplace transformation of  \eq{unravel}, with $\hat{\rho}^{(n)}(z)\equiv \int_0^\infty dt e^{-zt}  \rho^{(n)}(t) $ and similarly for all other quantities, 
\be\label{rhonconditioned}
\!\!\!\!
\hat{\rho}^{(n)}(z)\! &=&\! \sum_{ l_1...l_n}   \hat{S}_{l_n}(z) \kket{l_n} 
... \hat{w}_{l_3l_2}(z) \hat{w}_{l_2l_1}(z)  \hat{v}_{l_1}(z).
\ee
Due to the simple structure of \eq{rhonconditioned} in Laplace space, many of the Markovian waiting time calculations without feedback \cite{Bra08} carry over here.

\subsection{Moment generating function}
Counting statistics is introduced by counting fields  $\chi_{l}$ via phase factors $e^{i\chi_{l} }$ multiplying individual jump operators  $\mathcal{J}_{ll'}$. 
In the equation for the density operator $\hat{\rho}^{(n)}(z)$ after $n$ jumps, \eq{rhonconditioned}, the statistics of the jumps can then be obtained from the moment generating function $\hat{G}(\{\chi_l\},z)$. This function is defined as 
\be\label{Gdef}
\hat{G}(\{\chi_l\},z) &\equiv& \bbra{0}  \hat{S}(z) \kket{0} + 
\sum_{n=1}^\infty\sum_{l_1=1,...,l_n=1}^M  
\hat{u}_{l_n}(z)e^{i\chi_{l_n}} \nonumber\\
&...& \hat{w}_{l_3l_2}(z)  e^{i\chi_{l_2}}   \hat{w}_{l_2l_1}(z)  \hat{v}_{l_1}(z),
\ee
where 
\be\label{uldef}
\hat{u}_{l_n}(z)\equiv  \bbra{0}\hat{S}_{l_n}(z)\kket{l_n}. 
\ee
Note that we do not count the very first jump here - this would yield another factor $ e^{i\chi_{l_1}}$. 

Introducing the diagonal matrix  $e^{i\chi}\equiv$diag$(e^{i\chi_l})$, the sums of the $l_i$ in \eq{Gdef} define a product of matrices multiplied by the vectors $\hat{\vecu}^T(z)$, \eq{uldef} from the left and $\hat{\vecv}(z)$, \eq{vldef}, from the right, for example $\hat{\vecu}^T(z)   e^{i\chi}        \hat{\vecW}(z)      \hat{\vecv}(z)$ for $n=1$ with the matrix $(\hat{\vecW}(z))_{ll'} \equiv \hat{w}_{ll'}(z)$. The sum over $n$ in \eq{Gdef} now leads to the geometric series
\be\label{Gchiz}
\hat{G}(\{\chi_l\},z) &=& \bbra{0}  \hat{S}(z) \kket{0}  \nonumber\\
&+&  \hat{\vecu}^T(z) \left[ 1-  e^{i\chi}  \hat{\vecW}(z)     \right]^{-1}  \hat{\vecv}(z),
\ee
which generalizes the relation between counting statistics and waiting time distributions \cite{Bra08} to the feedback-controlled case.
The equality 
\be
\label{FCS_multiple}
\det\left[1- e^{i\chi}\mathbf{W}(z) \right]=0
\ee
defines a polynomial in $z$, of which the zero $z_0(\{\chi_k\})$ with $z_0(0)=0$ determines the  counting statistics in the long-time limit \cite{Bagrets}. Examples for this behaviour are provided below.

\subsection{Unidirectional process}\label{section_unidirectional}
For the unidirectional process \eq{pntunidirectional}, in the control scheme with conditioning upon the previous jump as in \eq{previous}, the rates $\gamma( t|\{t_{n}\})$ in \eq{waituni} have the specific form
\be\label{wfeedback}
\gamma( t|\{t_{n}\}) = \gamma(t-t_n),\quad t\ge t_n.
\ee
The equation of motion belonging to  this protocol is an integral equation
\be\label{pnintegral}
p(n,t) = \int_0^t dt'w(t-t') p({n-1},t') + \delta_{n,0}  p({0},t)
\ee
with $p(0,t)\equiv  e^{-\int_{0}^{t} dt'\gamma(t')}$, and where the memory kernel $w(t-t')$ is given by the waiting time distribution 
\be\label{wtau}
w(\tau)\equiv \gamma(\tau)  e^{-\int_{0}^{\tau} dt'\gamma(t')},
\ee
cf. \eq{waituni}.
For the feedback conditioned on the previous jump, we found no
simple expression in terms of  a second derivative of the idle time distribution \cite{Thomas13}, although it would
be interesting to explore such as relation, also for general feedback schemes.

The simple idea behind this type of control is to choose the rates such that a particular form of waiting times $w(\tau)$ is generated. The time--dependent modulation of the rates $\gamma(t-t_n)$ starts again and again  after the $n$-th jump occuring at times $t_n$. This renders the rates themselves as random quantities, in contrast to open loop control with a protocol  $\gamma(t|\{ t_n \}) = \gamma(t)$ independent of the $t_n$.

The probabilities $p(n,t)$ define the moment generating function
\be\label{Gunidef}
{G}(\chi,t) \equiv \sum_{n=0}^\infty   e^{i\chi n}{p}(n,t),
\ee
and its Laplace transformed $\hat{G}(\chi,z)$  corresponding to \eq{Gchiz},
now  only involves scalars, with $\hat{\vecv}(z) =  \hat{\vecW}(z) \equiv \hat{w}(z)$ and 
$\bra{0}  \hat{S}(z) \ket{0} = \hat{u}(z)=\hat{p}(0,z)$, where $\hat{p}(0,z)\equiv \int_0^\infty dt e^{-zt} e^{-\int_0^t dt' \gamma(t')}$. We thus have
\be\label{Gchiuni}
\hat{G}(\chi,z) 
&=&  \hat{p}(0,z) +  \hat{p}(0,z) \left[  e^{-i\chi} - \hat{w}(z)     \right]^{-1}  \hat{w}(z) \nonumber\\
&=& \frac{\hat{p}(0,z)}{1-  e^{i\chi}   \hat{w}(z)    }\\
\hat{w}(z) &\equiv&  \int_0^\infty dt e^{-zt}           \gamma(t)   e^{-\int_0^t dt' \gamma(t')},
\ee  
which we also obtain directly from \eq{pnintegral} with 
$
\hat{p}(n,z) = 
 [\hat{w}(z)]^n  \hat{p}(0,z) 
$ 
and by summing the geometric series of the Laplace transformed \eq{Gunidef}.

The probabilities $p(n,t)$ are therefore linked to the waiting time distribution $w(\tau)$  via the expression for the  moment generating function $\hat{G}(\chi,z)$, \eq{Gchiuni}. In the time domain, the long--time dynamics of $G(\chi,t)$ is determined by a zero $z_0(\chi)$ of the denominator $1-  e^{i\chi}   \hat{w}(z) $ in the moment generating function $\hat{G}(\chi,z)$ in Laplace space, \eq{Gchiuni}, with ${G}(\chi,t\to\infty) \sim \exp[t z_0(\chi) ]$ and $z_0(\chi=0)=0$ \cite{Bra08}. 

An important quantity in full counting statistics are cumulants of the distribution $p(n,t)$. The $k$--th cumulant is defined as 
\be\label{kthcumulant}
C_k(t) \equiv \left.\frac{\partial^k}{\partial (i\chi)^k} \ln G(\chi,t)\right|_{\chi=0},
\ee
and thus  $C_k(t\to \infty) = t  \left.\frac{\partial^k}{\partial (i\chi)^k} z_0(\chi)\right|_{\chi=0}$. Differentiating the denominator equation $1-  e^{i\chi}   \hat{w}(z_0(\chi))=0 $ twice with respect to $\chi$, we find
\be\label{Cwrelation}
\lim_{t\to \infty} \frac{C_2(t)}{C_1(t)} = \frac{\langle \tau^2 \rangle -  \langle \tau \rangle^2    }{\langle \tau \rangle^2} ,
\ee
where $\langle \tau^k \rangle$ is the $k$--th moment of the waiting time distribution $w(\tau)$. 
For feedback conditioned on the previous jump, \eq{Cwrelation} relates the full counting statistics directly to the width of the waiting time distribution as expressed by its variance var$(\tau)\equiv\langle \tau^2 \rangle -\langle \tau \rangle^2 $. This is completely analogous to the situation without feedback \cite{DHHW92,Bra08,Albert11}.

We note that in general, the result \eq{Cwrelation} does no longer hold for other feedback protocols. An example is the time-versus-number feedback with rates $\gamma(t,n) \equiv \gamma[ 1 + g (\gamma t - n)]$, \eq{FBexperiment} with $w(\tau)$ given by \eq{wtFBexp}. In this case, is has been shown \cite{Bra10} that the second cumulant converges towards a constant 
\be\label{C2freezing}
C_2(t\to \infty) = \frac{1}{2g},
\ee
whereas the first cumulant $C_1(t) = \gamma t$, and the ratio
$ \frac{C_2(t)}{C_1(t)}$ becomes zero at large times. As a consequence, the relation \eq{Cwrelation} is no longer valid for this feedback protocol.

\subsection{Gamma distribution example}\label{section_example}
A form of $\hat{w}(z)$ in Laplace space convenient for analytical treatment is 
\be\label{wzwcontrol}
\hat{w}(z)\equiv  \left(1+ \frac{z}{ (g+1) \gamma_0 }\right)^{-(g+1)},
\ee
with a feedback parameter $g$ that interpolates between  a usual  Poissonian process with waiting time $w(\tau) = \gamma_0 e^{-\gamma_0\tau}$ (no feedback, $g=0$), and 
the deterministic $w(\tau) =  \delta(\tau - \gamma_0^{-1})$ for $g\to \infty$ (where $\hat{w}(z) = e^{z/\gamma_0} $ in Laplace space \cite{Bra08}). 

\begin{figure}[t]
\centerline{\includegraphics[width=0.5\columnwidth]{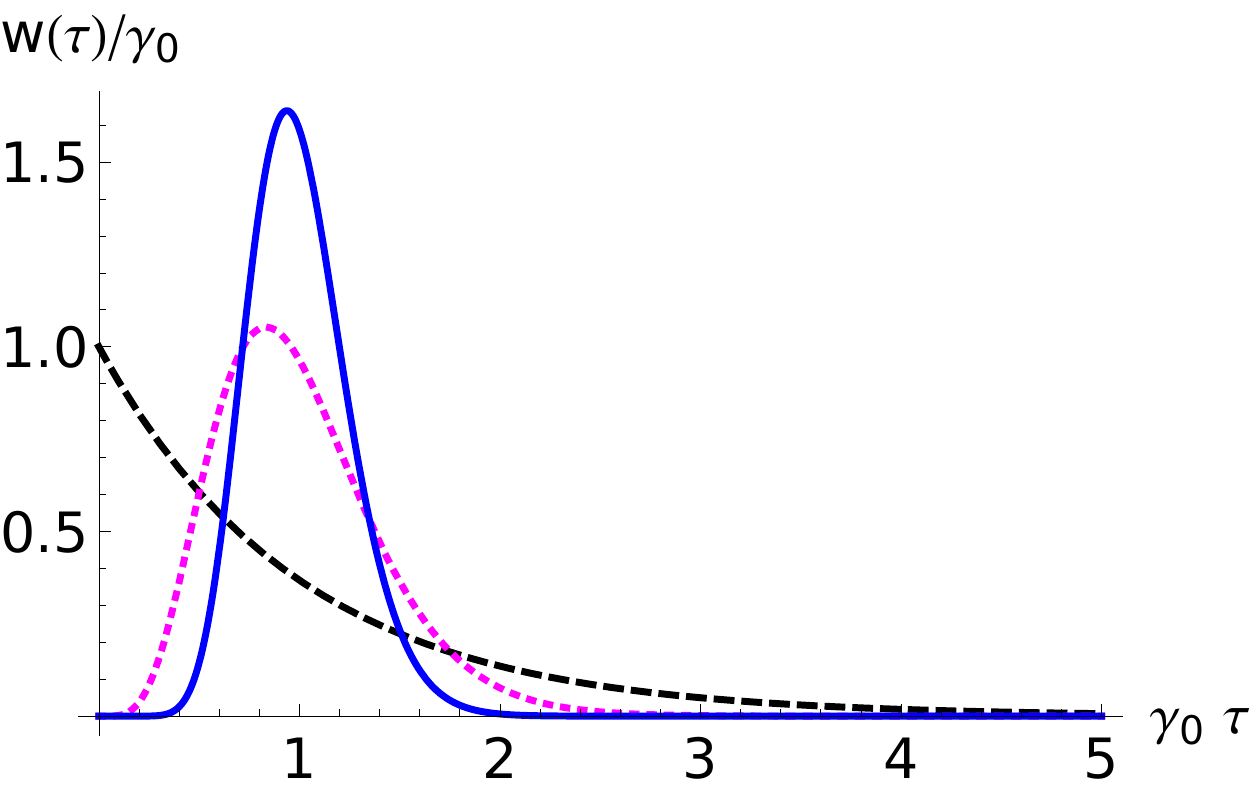} 
\includegraphics[width=0.5\columnwidth]{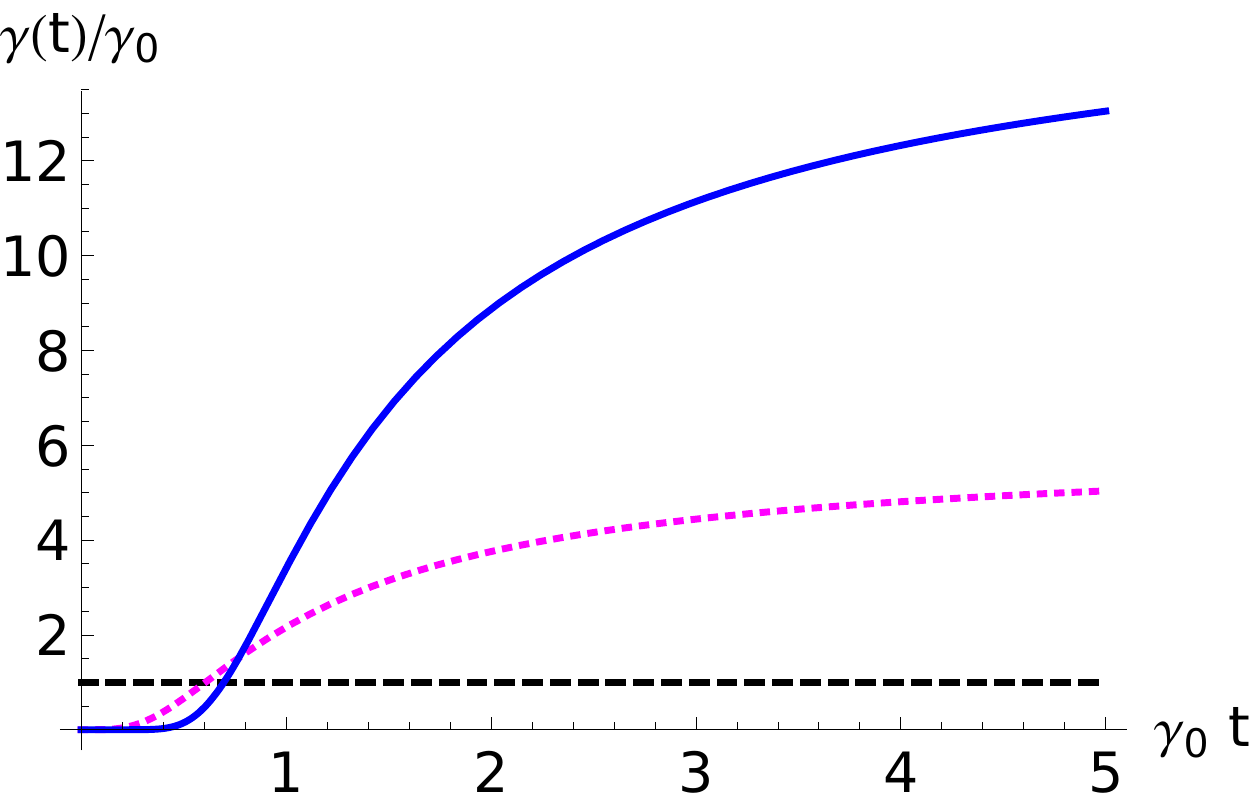} }
\caption[]{\label{figwgamma} Waiting time distributions $w(\tau)$, \eq{waituni}, (LEFT) of  single jump process with rates $\gamma(t)$ (RIGHT)  with a time-dependence according to \eq{rateuni} that starts again after each jump. Feedback strength $g=0$ (black, dashed), $g=5$ (magenta, dotted) and $g=15$ (blue, solid).}
\end{figure}

The $w(\tau) $ belonging to \eq{wzwcontrol} then have the form of a Gamma distribution, 
\be\label{waituniGamma}
w(\tau) = \frac{(g+1) \gamma_0 e^{-(g+1) \gamma_0 \tau}((g+1)\gamma_0 \tau)^g}{\Gamma(g+1)},
\ee
where $\Gamma(a)\equiv \int_0^{\infty} dt e^{-t} t^{a-1} $ is the Gamma--function. We obtain the protocol for the jump rate belonging to this waiting time distribution, i.e., the time dependence $\gamma(t)$,
via \eq{wtau}.
We use $w(\tau) = \gamma(\tau) p(0,\tau)$ and $\dot{p}(0,\tau)= -w(\tau) $ with ${p}(0,\tau)=1 - \int_0^{\tau} dt w(t)$, which leads to 
\be\label{rateuni}
\gamma(t) = \gamma_0 \frac{ (g+1) e^{-(g+1) \gamma_0 t}((g+1)t)^g}{\Gamma(g+1,(g+1) \gamma_0 t) }
\ee
with the incomplete Gamma--function  $\Gamma(a,x)\equiv \int_x^{\infty} dt e^{-t} t^{a-1} $.  
Fig. (\ref{figwgamma}) shows $w(\tau)$ and $\gamma(t)$ for three different values of the feedback strength $g$. 

The waiting time distribution $w(\tau)$, \eq{waituniGamma}, fulfills 
\be
\ew{\tau} = \frac{1}{\gamma_0},\quad  \mbox{\rm var}(\tau)=\frac{1}{\gamma_0^2(g+1)}.
\ee
In Laplace space, 
from  $\hat{p}(n,z) =  [\hat{w}(z)]^n  \hat{p}(0,z)$, $\hat{p}(0,z) = (1-\hat{w}(z))/z$ and \eq{wzwcontrol}, we obtain
the  probabilities $p(n,t)$ in the time-domain by Laplace inversion as 
\be\label{pnuni}
p(n,t) &=& \frac{\Gamma((g+1)(n+1),(g+1) \gamma_0 t)}{\Gamma((g+1)(n+1))} \nonumber\\
&-& \frac{\Gamma((g+1)n,(g+1) \gamma_0 t)}{\Gamma((g+1)n)}.
\ee
The $p(n,t) $ as a function of $n$ are very close to Gaussians 
\be\label{pnGauss}
p(n,t)\approx p^{\rm G}(n,t) \equiv \frac{1}{\sqrt{2\pi \gamma_0 t /(g+1)}}e^{-\frac{1}{2} \frac{(\gamma_0 t - n)^2}{\gamma_0 t /(g+1) }}
\ee
at times $\gamma_0 t \gg 1$, as we also checked numerically (not shown here). The long--time dynamics is determined by a zero $z_0(\chi)$ of the denominator $1-  e^{i\chi}   \hat{w}(z) $ in the moment generating function $\hat{G}(\chi,z)$ in Laplace space, \eq{Gchiuni}, with 
${G}(\chi,t\to\infty) \sim \exp[t z_0(\chi) ]$ and 
\be
z_0(\chi)=(g+1)\gamma_0 \left( e^{ i \frac{\chi}{g+1}}-1\right).
\ee
The first and second cumulants $C_1$ and $C_2$ at large times then simply follow by taking derivatives according to \eq{kthcumulant},
\be\label{C12}
C_1 \sim \gamma_0 t ,\quad C_2 \sim \frac{\gamma_0 t}{g+1}.
\ee
Feedback control conditioned on the previous jump thus reduces the width of the distribution $p_n(t)$ by a factor $g+1$, but leaves the first moment $C_1$ unchanged. 

\section{Optimized feedback control }\label{optimization}
In this section, we combine feedback conditioned on the previous jump with ideas from optimal control theory \cite{Kirk2004}.

\subsection{Optimization goal}
One goal of the feedback could be to generate a particularly regular series of (still stochastic) jumps, i.e., a waiting time distribution that approaches a delta function, $w(\tau) \to  \delta(\tau - \gamma_0^{-1})$, with jumps separated by regular time intervals denoted here as 
the inverse of a nominal rate $\gamma_0$.  In order to have a concrete example, we will use the particular form \eq{waituniGamma}
with feedback parameter $g$ for the 
unidirectional stochastic process from Sect. \ref{section_example}. As mentioned there, $g$ interpolates between  a usual  Poissonian process with waiting time $w(\tau) = \gamma_0 e^{-\gamma_0\tau}$ (no feedback, $g=0$), and the deterministic  $w(\tau) =  \delta(\tau - \gamma_0^{-1})$ for $g\to \infty$.

\subsection{The cost functional and its optimisation}
The feedback protocol \eq{wfeedback} gives the relationship between the waiting time distribution $w(t)$ and the feedback-controlled rate $\gamma(t)$, to be $w(t) = \gamma(t)  e^{-\int_{0}^{t} dt'\gamma(t')}$,
\eq{wtau}.
Inversion of this relation can be used to find a prescription for the rate $\gamma(t)$ required to provide a particular target waiting time distribution, $w_T(t)$. However, this does not guarantee that the rate is
realistic, or be one that could be implemented experimentally.

We now address this issue by introducing a cost function and optimisation procedure to find an optimal $\gamma(t)$ given the existence of these external considerations.
In the first step, we introduce a cost functional 
\be
  J  = J_w + a J_a + b J_b
  \label{EQ:Jtot}
  .
\ee
Here $J_w$ is a measure of the distance between the actual waiting time distribution and the target. We take the simplest choice of a quadratic cost function and write
\be
  J_w 
  = 
  \int_0^\infty
  dt \,
  \rb{w(t) - w_T(t)}^2
  \label{EQ:Jw}
  .
\ee
The second and third terms read 
\be
  J_a 
  = 
  \int_0^\infty
  dt \,
  \gamma^2(t)
  ;\quad\quad 
  J_b
  = 
  \int_0^\infty
  dt \,
  \dot{\gamma}^2(t)
  \label{EQ:Jab}
  ,
\ee
and account for constraints on the magnitude and rate-of-change of the rate respectively. Again, we choose a quadratic form for simplicity.
The parameters $a$ and $b$ describe the relative importance of these two considerations relative to the desire to match the waiting time distribution to its target. The problem is to find the rate $\gamma(t)$ that minimises the cost $J$ given a particular target waiting time distribution and set of parameters $a$ and $b$.

In experiments, rates are typically changed electrostatically by gate voltages: 
changing rates thus ultimately amounts to performing work on the system, and the speed of 
change in the rates corresponds to electrostatic power. 
In a realistic experiment, to determine such energetic costs reliably is probably  
quite difficult, as one would also have to consider the required  electronic circuits, memory etc. in the 
implementation of the feedback loop. Clearly, our phenomenological cost functionals are therefore  only
a crude approximation. 

The cost $J$ is a functional of the rate.  From \eq{wtau} we see that it depends not only on $\gamma(t)$ and $\dot\gamma(t)$, but also on the integral of $\gamma(t)$, which complicates matters. The scheme we employ to minimise $J$ is as follows. We first write \eq{EQ:Jtot} as
\be
  J  = 
  \int_0^\infty
  dt \,
  \left[
    \rb{\gamma(t) A(t) - w_T(t)}^2
    + a \gamma^2(t)
    + b \dot{\gamma}^2(t)
  \right]
  ,
\ee
with
$
  A(t) = e^{-\int_0^t dt' \, \gamma(t')}
$.
We then vary $J$ under the assumption that $A(t)$ is some {\em known} function of $t$, rather a functional of the unknown $\gamma(t)$.  With $A$ fixed, the variation of $J$ is straightforward, and gives the simple Euler-Lagrange equation
\be
  b \ddot{\gamma}
  - 
    \left[
      A^2(t) + a
    \right]\gamma(t) 
  + w_T(t) A(t) 
  = 0
  \label{EQ:EL}
,
\ee
subject to the natural boundary conditions
$\dot{\gamma}|_{t=0} = \dot{\gamma}|_{t=\infty} = 0$.  We then solve this problem iteratively.  First we calculate $A(t)$ using the known target rate without control $\gamma_T(t)$.  We then solve \eq{EQ:EL} equation for a new $\gamma(t)$ and use this to calculate a new $A(t)$. This procedure is then iterated until convergence is obtained.

\begin{figure}[tb]
  \begin{center}
     \includegraphics[width=0.49\columnwidth,clip=true]{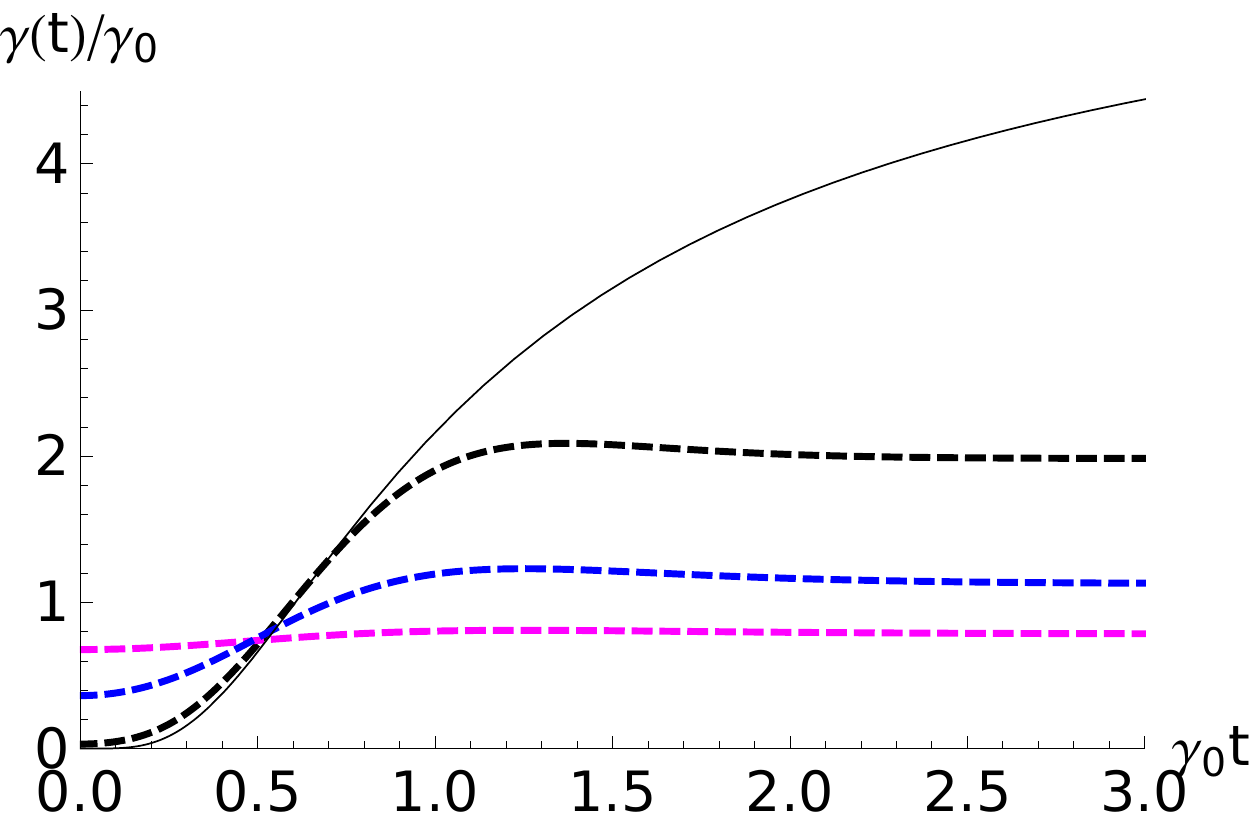}
     \includegraphics[width=0.49\columnwidth,clip=true]{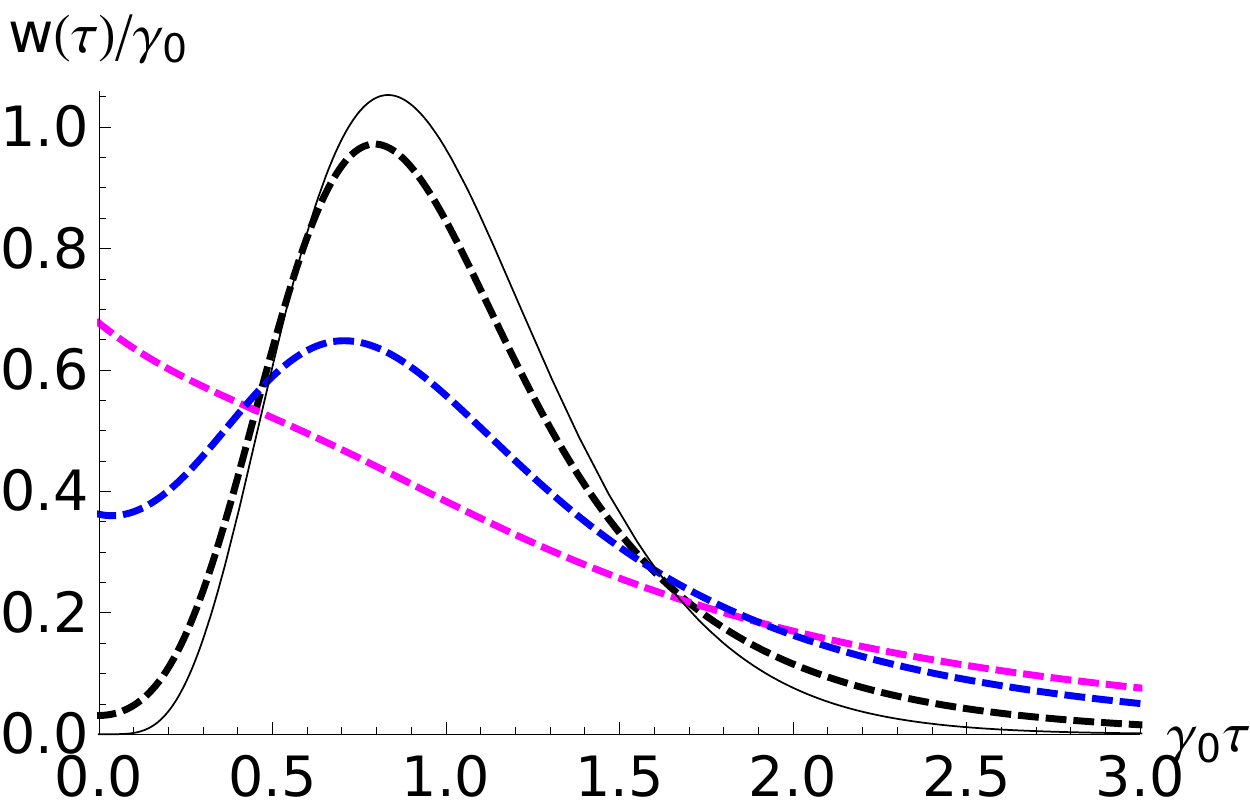}
  \end{center}
  \caption{
    Results from the optimisation of the cost functional $J$ with $a=0$, in which case we only impose a cost with the gradient of $\gamma(t)$. The target waiting time distribution is that in \eq{waituniGamma} with $g=5$.
    LEFT: the optimised rate function $\gamma(t)$. RIGHT: the corresponding waiting time distribution. The thin black lines show the target $\gamma(t)$ and $w(\tau)$, and optimised results are given for $b=0.01,0.1,1.0$ (dashed lines; black, blue, magenta).
    \label{FIG:a0}
  }
\end{figure}

\subsection{Optimization results}

\begin{figure}[tb]
  \begin{center}
      \includegraphics[width=0.49\columnwidth,clip=true]{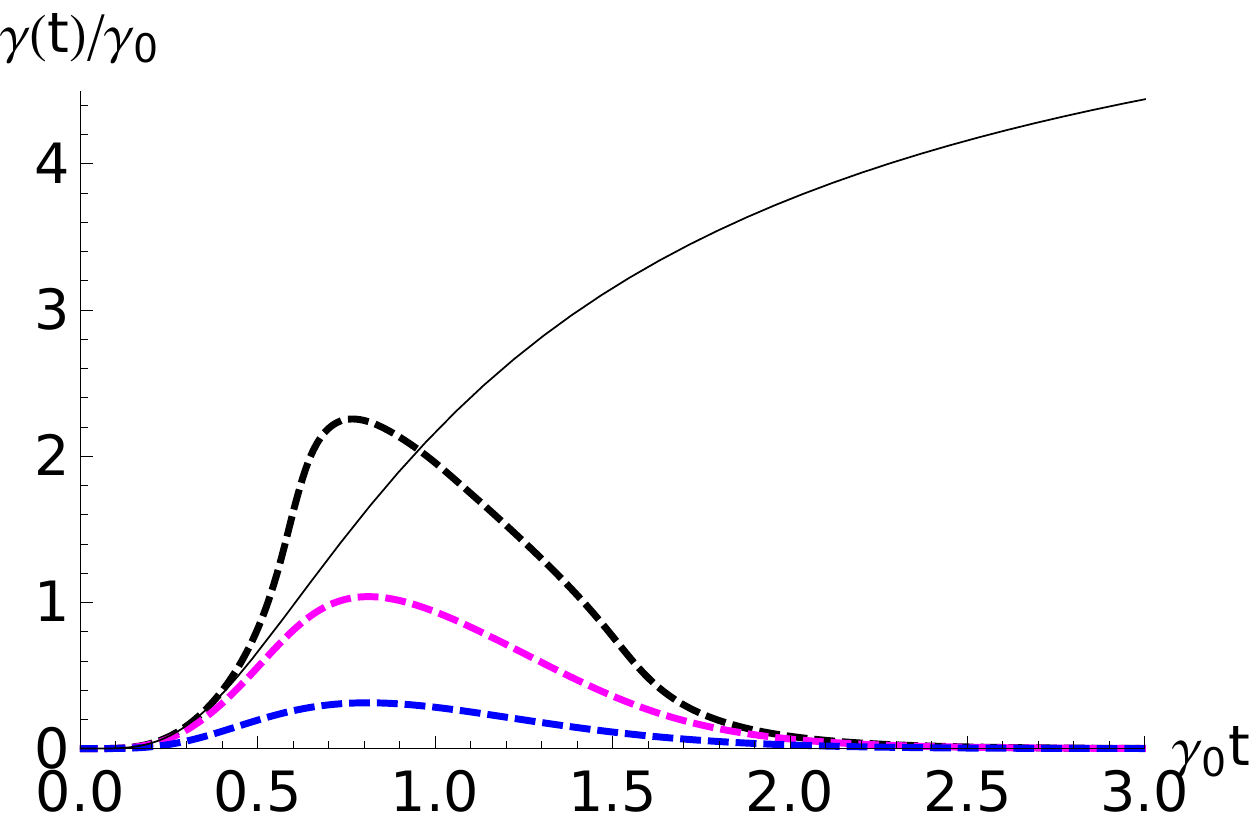}
     \includegraphics[width=0.49\columnwidth,clip=true]{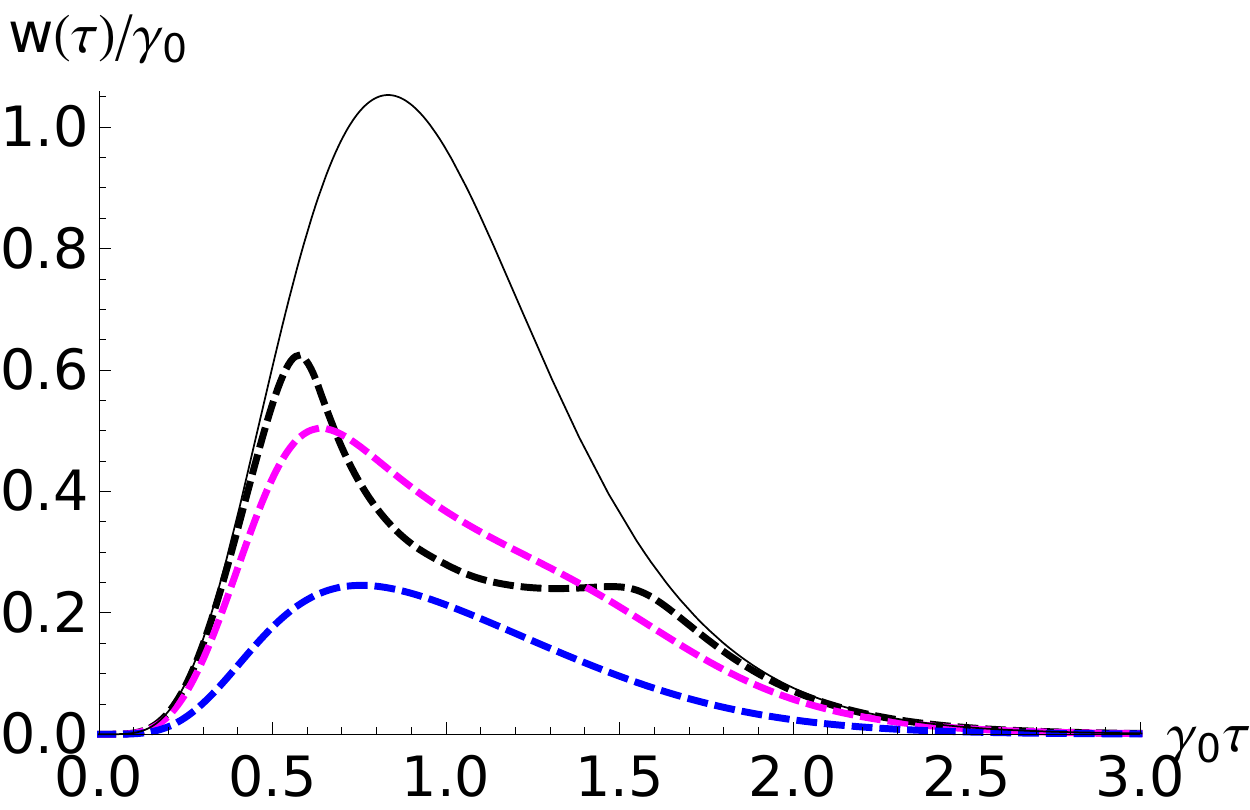}
  \end{center}
  \caption{
    As \fig{FIG:a0}, but here we set the rate parameter $b=0$ and investigate behaviour with cost parameter $a = 0.05,0.25,2.0$ (dashed lines; black, magenta, blue), i.e. we impose a cost associated with high rates.
    \label{FIG:b0}
  }
\end{figure}

We take as our target distribution that of \eq{waituniGamma} with $g=5$ and optimize $\gamma(t)$.  \fig{FIG:a0} and \fig{FIG:b0} show results for two end-point cases of this optimisation; the first with $a=0$ and the second with $b=0$.

The relative importance of $a$ versus $b$  will again depend on the details of the implementation via circuits for the 
gate voltage: it might require, e.g.,  additional resources to achieve fast switching rates reliably, or 
one is simply limited by certain upper bounds for the rates, as in the experiment \cite{FBexperiment16}.

%
With $a=0$, we associate a cost to the gradient of $\gamma(t)$.  Thus, in \fig{FIG:a0} we see a flattening of the optimised rate curve with increasing $b$. In large $b$ limit, the rate becomes flat and the waiting time distribution becomes Poissonian. Interestingly, even for small $b$, the rate at large times is significantly reduced by the optimisation.   This is because the high tail of the $\gamma_T(t)$ hardly effects the bulk of the waiting time distribution and can thus be culled by the optimisation.
With $b=0$, a finite value of $a$ makes minimization of the total area under the rate curve a priority.  Thus in \fig{FIG:b0}, we see an overall shrinkage in the rate with increasing $a$.  At large times, the value of the rate has little overall effect on the waiting time distribution, and thus this can be optimised away to zero.  
Interestingly, we see that as $a$ increases, the optimised waiting time distribution approaches the target $w_T(t)$ by first matching the front edge of the distribution, whilst still optimising for a suppressed tail at large time.

\section{Comparison of open and closed loop control}\label{section_comparison}
We now turn to a more detailed comparison of two feedback protocols in section \ref{sectionfp} in an analysis of open loop control versus 
feedback (closed loop) control conditioned on the previous jump. 
We define identical control goals in both schemes as the goal to achieve the same average waiting time $\ew{\tau}$ at a minimal width of the waiting time distribution. 
We quantify this goal by using the 
variance $\mbox {\rm var}(\tau) \equiv \langle \tau^2 \rangle-\langle \tau \rangle^2$ and introduce  the dimensionless Fano factor, cf. \eq{Cwrelation}, 
\be\label{Fdef}
F \equiv \frac{ \langle \tau^2 \rangle-\langle \tau \rangle^2}{\ew{\tau}^2}. 
\ee
Our reference will be the unidirectional Poissonian process \eq{pntunidirectional} without control, where $w(\tau) = \gamma_0 e^{-\gamma_0 \tau}$ with constant rate $\gamma_0$ and $F=1$.

\subsection{Piece-wise constant  rates}
For a simple comparison within an analytically tractable model we compare waiting time distributions with the controlled rates $\gamma( t_n|\{t_{n-1}\})$ given by piece-wise constant  rates $\gamma_0$ as in Fig.(\ref{costcompare}). For feedback (fb), we thus use rates
\be\label{gammafb}
\gamma_{\rm fb}(t)\equiv \gamma_0 \theta (t-\tau_0)
\ee
that are switched on at time $t=\tau_0$ after each  previous jump, while for open loop (ol) control we introduce periodic rates
\be\label{gammaol}
\gamma_{\rm ol}(t) \equiv \gamma_0 \sum_{n=-\infty}^\infty \chi_n(t),
\ee
which are continuously switched on and off with period $T$ and duration $\Delta T\ll T$. Here, the unit function $\chi_n(t) = 1$ in the time interval $t\in [nT, nT+\Delta T]$ and zero else. 
Further below, we will optimize the control in the sense of choosing between open or closed loop control based on the rates \eq{gammafb}, \eq{gammaol}.

In both protocols, these rates define a unidirectional  stochastic process \eq{pntunidirectional} with waiting time distributions \eq{waituni} controlled by $\gamma(t|\{t_n\})$. 
This process can be regarded as an  effective description of a two--barrier model as described in the following. 
The  model is defined by states $\ket{n,\sigma}$ and  $\ket{n,\sigma}$, $n\in \mathbb{N}$, where $\sigma=0,1$ denotes the 
state of a small region such as a quantum dot with a single level (`single electron transitor') that is  either occupied ($1$) or empty ($0$).
Particles enter one by one from a source reservoir at a fixed (un-controlled)  constant rate $\gamma_{\rm in}$ if the dot is empty, and leave at a (controlled) rate $\gamma(t|\{t_n\})$  into a drain reservoir
which is filled with $n$ particles, with $n$ increasing as time passes. 

The simplified description leading to the stochastic process $p(n,t)$, \eq{pntunidirectional}, 
then follow by  assuming a separation of time scales, in which the  whole stochastic process is goverened by the slow rate $\gamma(t|\{t_n\}) \ll  \gamma_{\rm in}$ only. 
For feedback control, the effective waiting time distribution 
\be\label{wfbset}
w_{\rm fb}(\tau) = \gamma_0 \theta (\tau-\tau_0) e^{-\gamma_0 (\tau-\tau_0)}. 
\ee
then follows from the convolution of subsequent waiting time distributions, i.e., the deterministic $w_{\rm in}(\tau)=\lim_{\gamma_{\rm in}\to \infty }\gamma_{\rm in} e^{-\tau \gamma_{\rm in}} =\delta(\tau)$ for jumps $0\to 1$, and  the  waiting time distributions for jumps $1\to 0$. 

In open-loop control, the limit $\gamma_{\rm in}\to \infty$ of immediate `re-charging' of the inner dot region after a jump $1\to 0$ would allow for arbitrarily small waiting times $\tau$ which would leave the control scheme {\em a priori} in disadvantage as compared with feedback. We exclude this by assuming a small delay for re-charging, 
which is assumed to occur, if the dot is empty,  deterministically (as in the feedback case)  at times $nT + \Delta T + 0^+ $ 
immediately after  the out-rate $\gamma_{\rm ol}$ is switched off, cf. Fig.(\ref{costcompare}), left. Counting the time between an in- and an
out-jump, with the relevant distribution denoted as $w_{\rm ol}(\tau)$, then always has to start 
at any of the times  $nT + \Delta T + 0^+ $. This assumption of deterministic re-charging simplifies the calculation drastically,
as one does not need to carry out an integration over the period $T$, and one can choose $n=0$ due to the time-translation invariance of the rate \eq{gammaol}. 
This leads to 
\be\label{wolset}
w_{\rm ol}(\tau) &\equiv& w_{\rm ol}(\tau+\Delta T + 0^{+} ,\Delta T + 0^{+})\\
&=& \gamma_0 \sum_{n=1}^{\infty}  \chi_n(\tau+\Delta T + 0^{+}) e^{-\gamma_0 \tau}  e^{-\gamma_0 (\Delta T - T )n}\nonumber
\ee 
as the relevant waiting time distribution.

\subsection{Costs of control schemes}
We now quantify the costs of the two control schemes using cost functionals as in section \ref{optimization}. 
For open loop control, we introduce the cost over one period as 
\be\label{Joldefinition}
J_{\rm ol} \equiv \frac{a}{T} \int_{0}^{T} dt \gamma_{\rm ol}^2(t) + ab \gamma_0^2   = \frac{a \gamma_0^2 \Delta T }{T} + ab \gamma_0^2,
\ee
where $a>0$ is a scale factor. Here, the first term in \eq{Joldefinition} corresponds to  the cost caused by the magnitude of the rate, and the second term $b \gamma_0^2 $ is a model for 
the cost of switching the rates on and off during one period $T$, determined by the parameter $b>0$ . We fix the scale factor using the fixed average $\ew{\tau}$ such that $a = \ew{\tau}^2$ 
and  the cost without feedback ($b=0$, $\Delta T = T$)  is unity, $J_{\rm ol} = \ew{\tau}^2\gamma_0^2=1  $. The dimensionless quantity \eq{Joldefinition} then defines the open loop cost in units of the no-control cost. 

For feedback control, the rates and thus the cost functional $ J_{\rm fb}(\tau)  \equiv             
\int_0^\tau dt \gamma_{\rm fb}^2(t)$ for an interval with waiting time $\tau$ becomes a stochastic quantity. A meaningful quantity for comparison with  open loop control, \eq{Joldefinition}, then involves  the 
average of   $J_{\rm fb} $ over an average waiting time $\ew{\tau}$ interval, and we define
\be\label{Jfbdefinition}
J_{\rm fb} &\equiv&   
\frac{a}{\ew{\tau} } \int_0^{\langle\tau\rangle}  d\tau     w_{\rm fb}(\tau) J_{\rm fb}(\tau) + a b \gamma_0^2 + c \nonumber\\
&=& \gamma_0 \ew{\tau}  + b (\gamma_0 \ew{\tau})^2 + c,
\ee
where we again used the same scaling factor  $a = \ew{\tau}^2$ and the switching cost, assuming the same parameter $b$ as in \eq{Joldefinition}  for simplicity. 
In addition, we have to consider the cost of continuously monitoring the system in order to start the feedback protocol right after each jump, a cost that is not present in the open loop scheme. Microscopic models for this kind of information related costs have been proposed in the context of 
Maxwell's demon and entropy flows recently \cite{SSEB2013,HE2014}. In our phenomenological model \eq{Jfbdefinition}, we therefore  postulate a fixed cost $c>0$ per interval $\ew{\tau}$ for the feedback observer.

\subsection{Results}
Using \eq{wfbset} and \eq{wolset}, we also obtain explicit expressions for the average waiting time $\ew{\tau}$, the variance and the Fano factor \eq{Fdef}.
For feedback control, these follow as 
\be
\langle \tau \rangle_{\rm fb} = \gamma_0^{-1} +  \tau_0,\quad F_{\rm fb}=\frac{1}{\left(\langle \tau \rangle_{\rm fb}\gamma_0\right)^{2}}.
\ee
For open-loop control, the expressions are lengthy but can be simplified in the limit $\gamma_0 T \gg 1$ and $T \gg \Delta T$ assumed in the following. The result is 
\be\label{ewtauol}
\ew{\tau}_{\rm ol} = \frac{T}{1- e^{-\alpha}}, \quad  
F_{\rm ol} = 1 - \frac{T}{\ew{\tau}_{\rm ol}},\quad \alpha \equiv \gamma_0 \Delta T,
\ee
where the parameter $\alpha$ defines the effective strength of the pulse \eq{gammaol}. Note that for  given  values of $\ew{\tau}_{\rm ol}$ and the Fano factor $F_{\rm ol}$,
the period $T$ is fixed, and thus  by \eq{ewtauol} so is the feedback strength $\alpha$  of the open-loop control scheme.

\begin{figure}[t]
\centerline{\includegraphics[width=0.5\columnwidth]{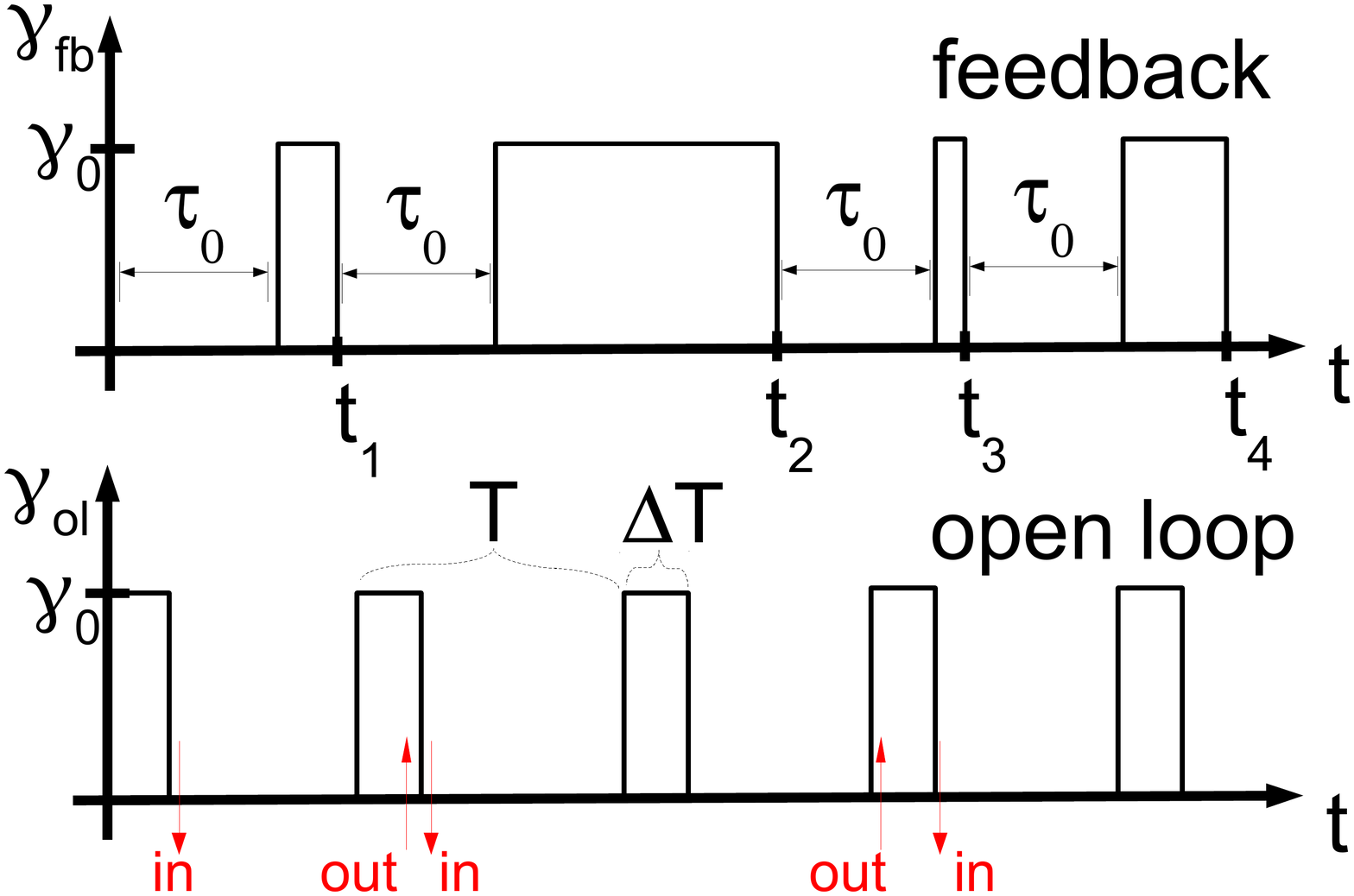} 
\includegraphics[width=0.5\columnwidth]{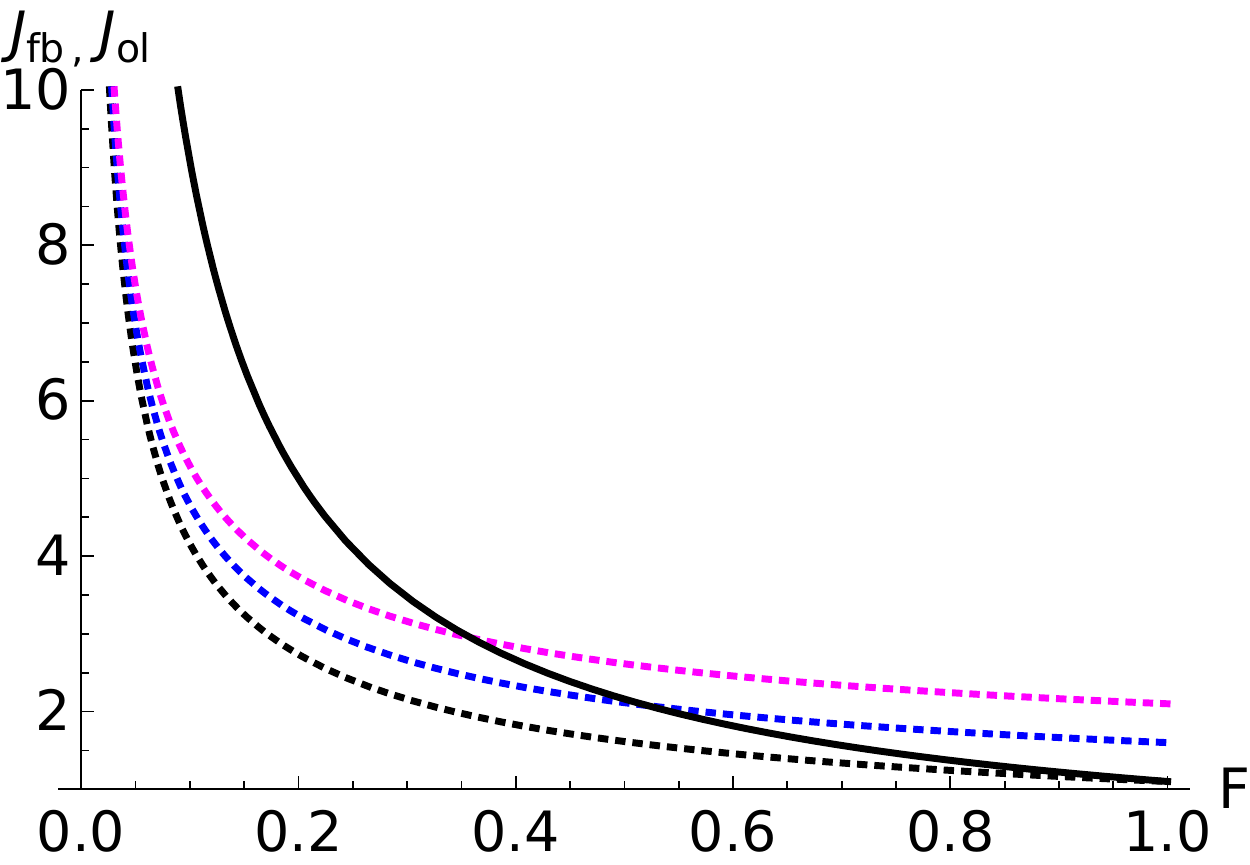}}
\caption[]{\label{costcompare} 
LEFT: Comparison of control schemes for piece-wise constant rates, \eq{gammafb}. Upper: feedback (closed loop) control starting after each jump at time $t_n$ with delay $\tau_0$. Lower: open loop control with rates \eq{gammaol} switched on during time intervals $\Delta T$ with period $T$. Red arrows show a possible charging and de-charging sequence in the two--barrier model explained in the text.
RIGHT: Comparision of costs for feedback control (dotted curves) $J_{\rm fb}$, \eq{Jfbdefinition}, with costs for open loop control (solid curve),  $J_{\rm ol}$, \eq{Joldefinition} as a function of scaled Fano factor $F$, \eq{Fdef}. Parameters for the observation cost $c$ in $J_{\rm fb}$ increase from bottom to top curve with $c=0$ (black), $c=0.5$ (blue), $c=1$ (magenta). Switch cost parameter $b=0.1$.

}
\end{figure} 

We are now in a position to carry out a quantitative comparison. 
First, in the extreme case of infinite cost with $\gamma_0,\alpha \to \infty$, not so surprisingly the waiting time distributions  becomes sharp in both schemes, $w(\tau) = \delta(\tau -\ew{\tau})$, with  open-loop period $T=\ew{\tau}$.
At any finite $\gamma_0$, 
for fixed Fano factor  $F_{\rm fb}= F_{\rm ol} =F$ as the control target, the comparison then  amounts to directly comparing the control costs in both schemes. 

The result of this comparison is shown in Fig. (\ref{costcompare}), where we plot the costs as a function of the Fano factor, which  are given by 
$J_{\rm fb}  = \frac{1}{\sqrt{F}} + \frac{b}{F}+ c$  and $J_{\rm ol} =   - [\log(F)]/(\sqrt{F} (1-F))  + \frac{b}{F} $. 
Clearly, at small enough Fano factor $F$, feedback control is always less expensive and thus  superior to  open loop control:  it is more efficient to achieve sharply peaked waiting time distributions
with feedback than with simple periodic driving. This advantage is particularly pronounced at small switching costs $b$ and small observation costs $c$.  
On the other hand, if the feedback goal is less ambitious and only a small reduction of $F$ from the non-controlled value $F=1$ is desired, the  observation costs $c$ are too high to make feedback efficient, and open loop control becomes the better choice.

\section{Thermodynamics of waiting time feedback control }\label{section_thermo}
In this last section, we make an attempt towards  analysing waiting time feedback control from a thermodynamic point of view.  In  other feedback schemes, this has been successfully achieved recently.
An example is the bi-partite splitting of a physical system into a controller and the controlled part, a situation where one can use the concept of mutual information and the flow of entropies. One can then interpret, e.g., devices that are close to -- within certain limits of parameters -- thermodynamic feedback paradigms such as Maxwell's demon \cite{SEKB11,AMP2011,SSEB2013,MNV2009}.

\subsection{Fluctuation relation}\label{section_bidirectional}
Our first observation concerns the role of detailed balance and a possible modification of the exchange fluctuation relations \cite{EHM2009}  in presence of waiting time feedback. 
In fact, the  protocol \eq{previous}  conditioned on the previous jump allows one to perform feedback control without modifying the exchange fluctuation relation
\be\label{exfr}
\lim_{t\to \infty} \frac{p(n,t)}{p(-n,t)} = e^{\mathcal{A}n}
\ee
for the bidirectional stochastic process \eq{bidirectional}. Here, $\mathcal{A}\equiv \beta (\mu_1-\mu_2)$ denotes the affinity in a situation with transport between two reservoirs at equal inverse temperature $\beta$ and chemical potentials $\mu_{1,2}$. 

We demonstrate \eq{exfr} for the particular example of a single tunnel junction between two fermionic reservoirs $1$ and $2$, with rates for forward ($1\to 2$) and backwards ($2\to 1$) jumps
\be\label{gammapm}
\gamma_+(t-t_n) &=&  \Gamma_+(t-t_n) f_1 ( 1-f_2)\nonumber \\
\gamma_-(t-t_n) &=&  \Gamma_-(t-t_n) f_2 ( 1-f_1). 
\ee
Here, $\Gamma_\pm(t-t_n)$ denotes the  bare, feedback controlled  tunnel rates without the thermal Fermi distributions $f_\alpha\equiv (e^{\beta(\epsilon-\mu_\alpha) }+1)^{-1} $. 

The crucial point now is that these rates fulfull detailed belance, as long as the bare rates $\Gamma_\pm(t-t_n)= \Gamma(t-t_n)$ do not depend on the direction ($\pm $) of the jump,
\be\label{detailedbalance}
\gamma_+(t-t_n) =  e^{\mathcal{A}}  \gamma_-(t-t_n).
\ee
In this situation, feedback control is still present in the form of conditioning the bare rates on the previous jumps and thereby modifying the waiting time distribution
and the full counting statistics $p(n,t)$, cf. \eq{Cwrelation}. However, the feedback is not sensitive to the direction of transport,  and thus in particular does not act like the Maxwell demon type rectifyer that would lead, e.g.,  to directional transport even at zero affinity $\mathcal{A}=0$ as in \cite{SEKB11,ES2012}.

This is corroborated by elevating the detailed belance condition of the rates, \eq{detailedbalance}, onto the fluctuation relation \eq{exfr} which here follows from a simple analysis of a symmetry in the
moment generating function $G(\chi,t)$: the two types of jumps lead to a $2\times 2$ waiting times matrix $\hat{\vecW}(z)$, \eq{Gchiz},  with elements 
$\hat{w}_{++}(z)=\hat{w}_{+-}(z)\equiv \hat{w}_{+}(z)$ and $\hat{w}_{-+}(z)=\hat{w}_{--}(z)\equiv \hat{w}_{-}(z)$ as the Laplace transforms of
\be\label{waitingbi}
w_{\pm}(\tau) = \gamma_\pm(\tau) e^{-\int_0^\tau dt' \left[  \gamma_+(t') +  \gamma_-(t')\right]}.
\ee
The long--time dynamics of the moment generating function $G(\chi,t)$ now follows from the determinant condition  \eq{FCS_multiple}, $\det\left[1-e^{i\chi}\mathbf{W}(z) \right]= 0$, with  the counting field matrix $e^{i\chi}=$diag$(e^{i\chi}, e^{-i\chi})$. This condition is invariant under the change $\chi\to -\chi + i\mathcal{A}$ owing to \eq{detailedbalance} and \eq{waitingbi}, and from the symmetry $G(\chi,t)= G(-\chi+i\mathcal{A},t)$ we obtain \eq{exfr} as usual \cite{EHM2009}.

\begin{figure}[t]
\centerline{\includegraphics[width=0.5\columnwidth]{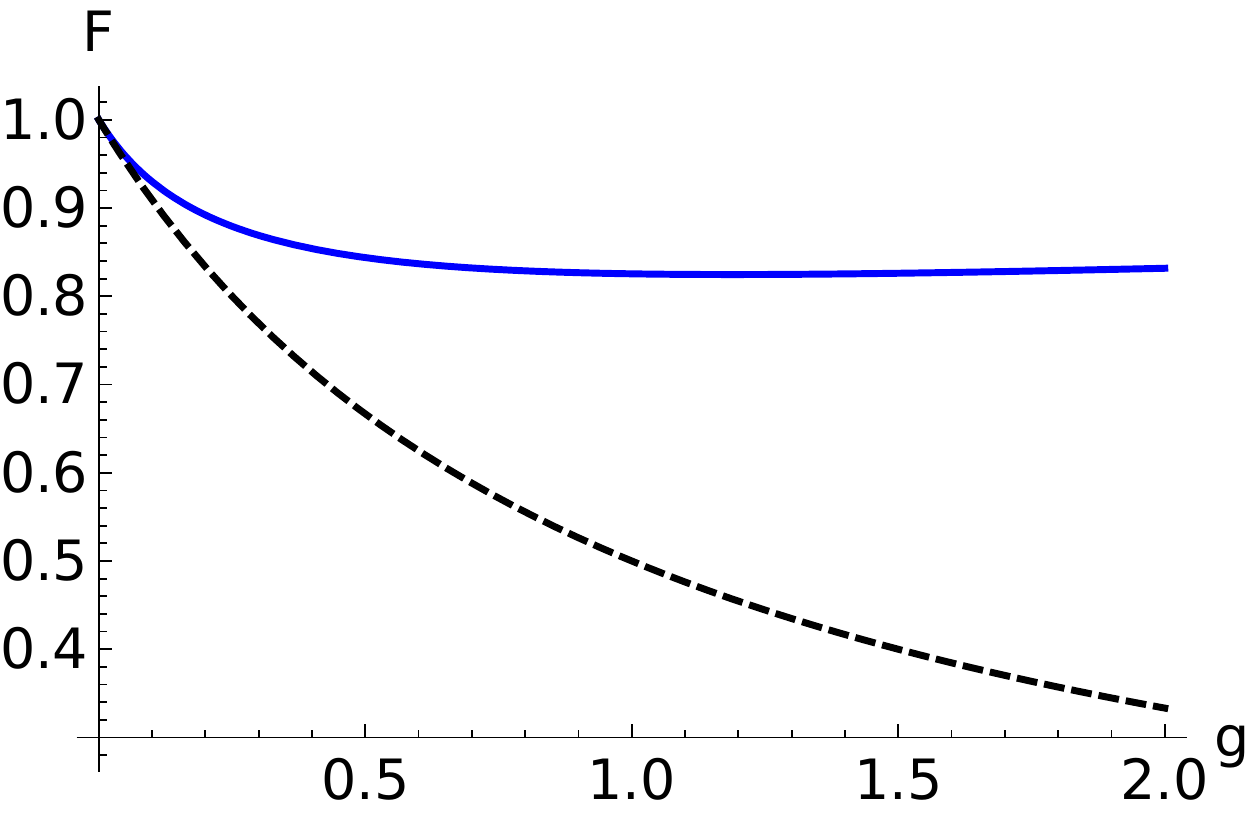} 
\includegraphics[width=0.5\columnwidth]{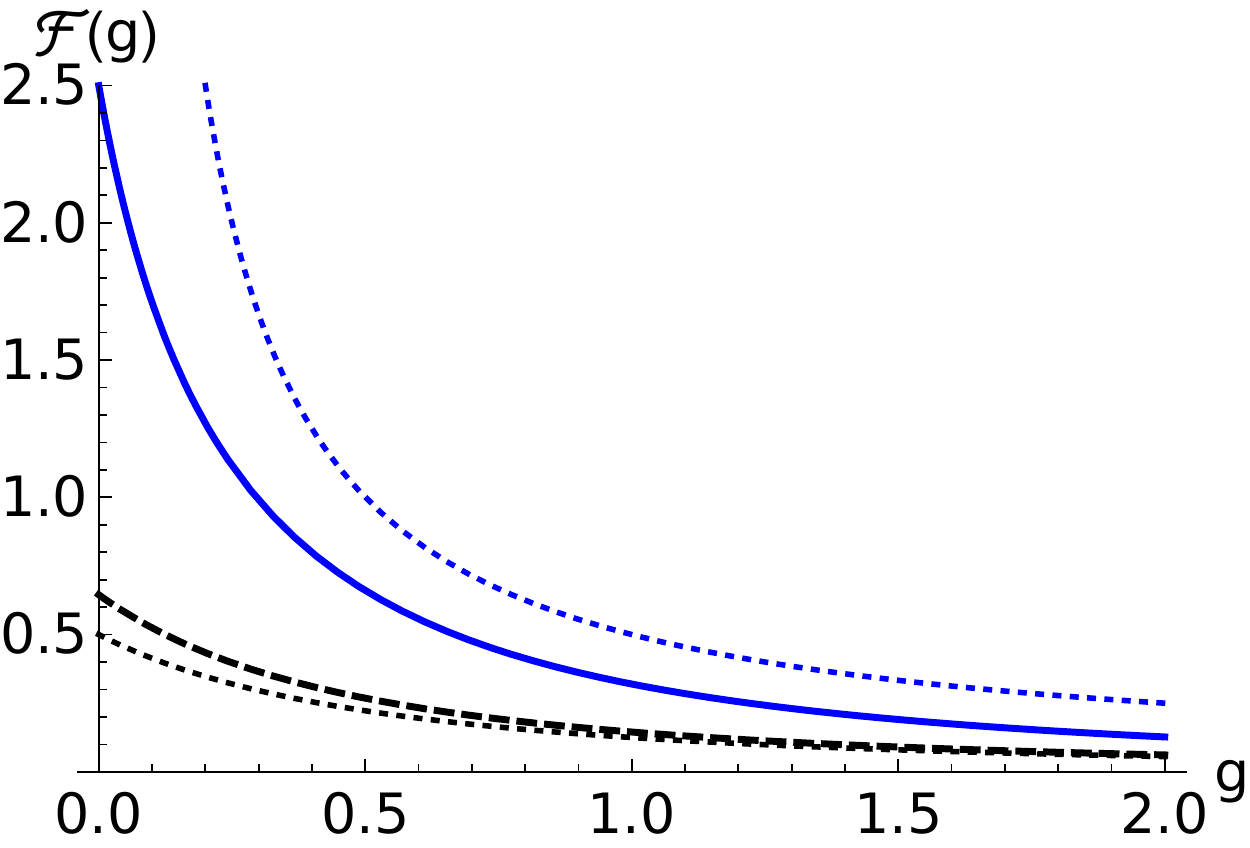}}
\caption[]{\label{Fishercompare}   Comparison of time-versus-number feedback (blue solid line), \eq{wtFBexp}, and feedback conditioned on previous jump (dashed black line), \eq{waituniGamma}, with dimensionless control parameter $g$. LEFT: Fano factors $F\equiv \frac{ \langle \tau^2 \rangle-\langle \tau \rangle^2}{\ew{\tau}^2}$. RIGHT:  Fisher information $\mathcal{F}(g)$ for waiting times, \eq{Dwdef}, and $\mathcal{F}_p(g)$ (dotted lines) for $p(n,t)$ in Gaussian approximation (see text).
}
\end{figure} 

\subsection{Information gain}
We can quantify the information gain in a phenomenological way by introducing  the Kullback--Leibler divergence (or relative entropies) between waiting time distributions $w_{g}(\tau)$ with different control parameters $g$,
\be\label{Dwdef}
D(g+\delta g,g)&\equiv& \int_0^\infty d\tau w_{g+\delta g}(\tau) \log \frac{ w_{g+\delta g}(\tau)}{ w_{g}(\tau)}\nonumber\\
&=& \frac{1}{2}\mathcal{F}(g) \delta g^2  + O(\delta g^3 ),
\ee
where the second line defines the Fisher information $\mathcal{F}(g)$. We use these quantities to compare the two feedback protocols;
feedback conditioned on the previous jump in the example of the Gamma distribution, \eq{waituniGamma}, and time-versus-number feedback, \eq{wtFBexp} and \eq{wntfull}. 
Note that this comparison is formal in the sense that the dimensionless feedback parameter $g$ will correspond to different physical parameters in any real implementation for each of the two schemes.

Fig. (\ref{Fishercompare}) shows the Fano factor $F$, \eq{Fdef} for the two waiting time distributions. In both cases, the Fano factor $F= 1- g+O(g^2)$ for small $g$, but the time-versus-number feedback clearly has larger Fano factors for larger values of $g$. 
For feedback conditioned on the previous jump, we find 
\be \label{Dwresult} 
D(g,0) = \log (g+1) - \log \Gamma(g+1) + g\Psi(g+1) -g,
\ee
where $\Psi( )$ denotes the Digamma function, and correspondingly $\mathcal{F}(0)=  \pi^2/6-1$.
Using the equivalence \eq{Cwrelation}  between Fano factors of $w_g(\tau)$ and the full counting statistics $p_g(n,t)$, we can introduce a corresponding  Kullback--Leibler divergence
\be\label{DSdef}
D_p(g+\delta g,g)\equiv \lim_{t\to\infty}\sum_{n=0}^\infty p_{g+\delta g}(n,t) \log \frac{ p_{g+\delta g}(n,t)}{ p_{g}(n,t)},
\ee
which we evaluate within the Gaussian approximation \eq{pnGauss} and which leads to a corresponding Fisher information $\mathcal{F}_p(g)= \frac{1}{2(g+1)^2}$. The good agreement between the latter and the 
Fisher information for $w_g(\tau)$  proves the quality of the Gaussian approximation for not too small $g$. 

In contrast, in the time-versus-number feedback protocol, the feedback--frozen second cumulant $C_2(t\to \infty) = \frac{1}{2g}$, \eq{C2freezing}, diverges for $g\to 0$ and so does the Gaussian approximation to the Fisher information, $\mathcal{F}_p(g)= \frac{1}{2g^2}$, in that case.  Since the close connection between $w(\tau)$ and $p(n,t)$, \eq{Cwrelation}, no longer holds, the  Fisher information  $\mathcal{F}(g)$ for the waiting times remains finite and is monotonously decreasing, with $\mathcal{F}(0)=\frac{5}{2}$.


\subsection{`Hardwiring' of a passive control system}\label{section_passive}
Finally and in the spirit of previous work on feedback in transport \cite{SSEB2013,Bra15}, we now present an analysis of a microscopic model which in its (passive) feedback operations is equivalent to a given active feedback scheme. 
As an example, we here consider again the unidirectional process with (actively) feedback--controlled  waiting time distributions given by the Gamma distribution \eq{waituniGamma} and 
$\hat{w}(z)\equiv  \left(1+ \frac{z}{ (g+1) \gamma_0 }\right)^{-(g+1)}$ in Laplace space, \eq{wzwcontrol}, where $g\ge 0$ is the feedback parameter.

In order to introduce passive control, we replace the single jump process by  a sequence of $N+1$ transitions $0\to 1 \to 2 \to ... \to N \to 0$ (a ring in the space of states). 
A concrete example for a physical realisation is as follows: a single electron jumps into an empty quantum dot (transition $0\to 1$) and then cascades down to lower energy levels in the dot ($1 \to 2 \to ... \to N$) due to emission of phonons until it leaves the dot from the lowest level ($N \to 0$) and the process starts afresh, cf. Fig. (\ref{waitingtime_fbscheme}).

For simplicity, we assume all jumps to occur at the same (inelastic phonon emission) rate $\gamma_N$, which results in a simple $(N+1)\times (N+1)$ matrix for the 
Liouvillian $\mathcal{L}(\chi,\eta)$ that includes a counting field $\chi$ for the electron (number of transitions $N\to 0$) and counting fields $\eta$ for the phonons (for each jump).
The density matrix (including counting fields) fulfills $\dot{\rho}(\chi,\eta,t) = \mathcal{L}(\chi,\eta) {\rho}(\chi,\eta,t) $. 

Here, we are interested in a reduced description where the internal levels remain unobserved and the corresponding counting field is set to zero, $\eta=0$. The counting is then for the electrons  (transitions $N\to 0$) only and determined by the moment generating function 
\be\label{Gelectrons}
\hat{G}(\chi,\eta=0,z) = \frac{1}{z}\frac{1-\hat{w}_N(z)}{1-e^{i\chi}\hat{w}_N(z) },
\ee 
with the waiting time distribution in Laplace space,
\be\label{waithard}
\hat{w}_N(z)  \equiv \left(1+\frac{z}{\gamma_N}\right)^{-(N+1)}.
\ee
These expressions are derived from the explicit matrix form of $\mathcal{L}(\chi,\eta)$ in Appendix B. 
Upon comparison we realize that \eq{waithard} is (by construction, of course), identical with the active feedback control form  \eq{wzwcontrol}, if we identify the feedback parameter $g$ there with $g=N$, the number of internal levels in our passive control scheme, and 
the rates  
\be\label{ratesscaling}
\gamma_N = (N+1) \gamma_0.
\ee
The case $g=N=0$ formally corresponds to no feedback and a single jump process (single tunnel barrier) only. 
We also note that $\hat{G}(\chi,\eta=0,z)$ then co--incides with the expression \eq{Gchiuni} for active control. 

We thus have identified a system where active feedback control can be simulated via passive control: 
here, active feedback control appears  in the reduced sub-system (electron) dynamics of a larger total system (electrons, phonons, dot).  

\subsection{Entropies in the passive control scheme}
We are now in a position to evaluate the entropic costs of our electron--phonon feedback system. For this purpose, 
we evaluate the Shannon entropies 
\be
S(t)\equiv -\sum_n p(n,t)\log p(n,t)
\ee
of the various probability distributions: the full counting statistics $p_{\rm el (ph)}(n,t), n\in \mathbb{N}$ of the electrons (emitted phonons), and the occupations $p_{\rm dot}(n), n=0,...,N$ of the dot levels. 

Clearly, in the long time limit one has $p_{\rm dot}(n)=\frac{1}{N+1}$ regardless of the state $n$ of the dot. On the other hand, 
the long--time behavior of the cumulant generating function for electrons and phonons is 
\be\label{Gchietalongtime}
\log {G}(\chi,\eta,t\to \infty) \sim t \gamma_N \left( e^{\frac{i\chi}{N+1}+i\eta}-1 \right),
\ee
cf. Appendix B, from which the second cumulants of the electron (el) and phonon (ph) statistics follow by differentiating twice and using \eq{ratesscaling},
\be\label{elphcumulants}
C_{2,{\rm el}}(t\to \infty)\sim \frac{\gamma_0 t}{N+1},\quad C_{2,{\rm ph}}(t\to \infty) \sim \gamma_0 t (N+1).
\ee
Note that the counting field $\chi$ (electron), this  co--incides with \eq{C12} of the active scheme as must be.

Using the Gaussian approximation as in \eq{pnGauss} with $n$ as a continuous variable, the cumulants \eq{elphcumulants} yield the Shannon entropies ($t \to \infty$) 
\be\label{entwire}
S_{\rm ph}^N(t) &=& \frac{1}{2} \ln ({2\pi \gamma_0 t}(N+1)) + \frac{1}{2}\nonumber\\
S_{\rm el}^N(t)& =&\frac{1}{2} \ln \left(\frac{2\pi \gamma_0 t}{N+1}\right) + \frac{1}{2} \nonumber\\
S_{\rm dot}^N&=&  \ln (N+1).
\ee 
As expected, $S_{\rm ph}^N(t) $ increases as a function of time $t$, and at a rate $(N+1)^2$ times faster than the entropy $S_{\rm el}^N(t)$ of the electrons. 
We interpret the two entropies  $S_{\rm ph}^N(t) $ and  $S_{\rm el}^N(t)$ as Shannon entropies characterizing the inner state of the electron and phonon counting devices, cf. Fig. (\ref{waitingtime_fbscheme}). 
Note that the (unphysical) divergence at time $t=0$ in \eq{entwire} is due to  the two delta-peak type initial conditions in the Gaussian approximation  \eq{pnGauss}.

At long times, phonon and electron entropies are balanced as
$
S_{\rm ph}^N(t) = S_{\rm el}^N(t) + S_{\rm dot}^N.
$    
If we consider the entropies \eq{entwire}  as a function of $N$, the thermodynamic costs of the  passive feedback mechanism become clear: increasing the number of dot levels $N$ logarithmically suppresses the electronic FCS entropy  $S_{\rm el}^N(t)$, thus making the electronic transport more regular with a sharper waiting time distribution, cf. \eq{Cwrelation}. This occurs 
at the expense of logarithmically increasing the entropy in the `feedback device', i.e. by increasing the entropy of the dot and generating more and more phonons. 

Finally, we can use the expressions \eq{entwire} and quantify the relative costs of this passive feedback scheme in terms of a (phenomenological)  efficiency
\be
\eta \equiv \frac{- \Delta S_{\rm target}}{\Delta S_{\rm resource} },
\ee
i.e., the ratio of decrease  of entropy of the target sub-system (the electrons leaving the system, $\Delta S_{\rm target}= S_{\rm el}^N(t) -S_{\rm el}^0(t)$) and the cost in terms of an increase in entropy of the passive controller providing the resource of the feedback control (the dot and the phonons, $\Delta S_{\rm resource} =  S_{\rm ph}^N(t)-S_{\rm ph}^0(t)+  S_{\rm dot}^N-  S_{\rm dot}^0$). Using 
\eq{entwire}, we find the value $\eta=\frac{1}{3}$ independent of the number $N>0$.

In contrast to our previous `hardwiring' scheme of Maxwell demon feedback \cite{SSEB2013,Pekola15}, the  passive feedback scheme in Sect. \ref{section_passive} unfortunately appears less relevant for a direct experimental realisation. In the analogon of the active feedback protocol \eq{wzwcontrol}, the stochasticity of the time intervals between the jumps into the drain reservoir is simply reduced by  `brute force', i.e., many copies of the same stochastic process in series, with the rates scaling up, \eq{ratesscaling}. This ultimately amounts to an `over--scaled' version of the law of large numbers where due to scaling the variance $C_{2,{\rm el}}$ vanishes in the limit $N\to \infty$, cf. \eq{elphcumulants}, and the whole process becomes deterministic.  

\section{Conclusion}  

Our results demonstrate that active feedback control of waiting times can be formulated in a natural way within the master equation formalism. Out of the 
numerous possible feedback protocols based on trajectories of measurement results, we only considered two (time-versus-number and feedback conditioned 
on the previous jump) in greater detail. Our models indicate that relatively large feedback strengths are required to achieve a noticeable control over the waiting times,
in contrast to, e.g., the time-versus-number feedback control of the full counting statistics   \cite{Bra10,FBexperiment16} where already small feedback yields a large control effect.
In general and in agreement with intuition, the control of short-time fluctuation thus appears to be  much harder than the control of the statistics over long times.

We also showed that active feedback protocols can be improved  by methods from optimized control theory. We believe this to be a new 
direction in the field of quantum feedback control. Here, further routes to explore might be various kinds of `online' optimization schemes that 
work with individual trajectories (measurement records) and not with {\em a priori} fixed feedback protocols.

A desirable extension into the passive feedback direction would  be a smarter and yet simple physical realisation of waiting time feedback, possibly using systems involving quantum coherences such as in the example \eq{Rabi} of resonance fluorescence in Sect. \ref{section_examples}. Recent work in this direction is, e.g.,  
the analysis \cite{Schulze2014} of  photon bunching as observed \cite{Reitzenstein11} in a quantum-dot laser with optical feedback, based on a microscopic quantum model for the $g^{(2)}$--correlation function.

At the same time, we argue that the decision between active and passive control  has to be made case by case depending on the feedback goal and the corresponding feedback protocol. For example, in the `number-versus-time' protocol \cite{Bra10,FBexperiment16} discussed above, active feedback has been proven to be very successful experimentally. In contrast, the passive feedback version of that protocol  \cite{Bra15} involves relative complicated interactions among particles that are unrealistic for, e.g., electronic systems. 
From such a perspective, and also in view of our results on optimization of waiting time feedback in sections \ref{optimization} and \ref{section_comparison}, the active schemes look quite promising.

What needs further clarification then, though, remains an understanding of this kind of feedback from a thermodynamical perspective beyond a mere comparison of phenomenological costs and efficiencies.

\section*{Acknowledgements}
We thank T. Wagner and R. J. Haug for showing us data on waiting time distributions for their experiment \cite{FBexperiment16}.
We also thank  G. Schaller and P. Strasberg for valuable discussions, and  acknowledge support by the DFG  via projects BR 1528/7-1, 1528/8-1, 1528/9-1, SFB 910 and GRK 1558.

\begin{appendix}
\section{Waiting times for time-versus-number feedback}
Here, we derive the  waiting time distribution $w(\tau)$, \eq{wtnscheme} for the time-versus-number feedback protocol $\gamma(t,n) \equiv \gamma[ 1 + g (\gamma t - n)]$,  \eq{FBexperiment}, of the unidirectional stochastic process $p(n,t)$, \eq{pntunidirectional},
\be\label{pnFBexperiment}
p_n(t) &=& \int_0^{t} dt_n ...\int_0^{t_{2}}dt_{1} e^{-\int_{t_n}^t dt'\gamma(t',n) } \times  \\
&\times& 
w_n(t_n,t_{n-1}) ...w_2(t_2,t_{1}) w_1(t_1,0), \nonumber
\ee
with  waiting time distributions $w_n(t_n,t_{n-1}) \equiv \gamma(t_n,n-1) e^{-\int_{t_{n-1}}^{t_n} dt' \gamma(t',n-1)}$, \eq{wndef}.  
\eq{pnFBexperiment} corresponds to the simple master equation ($n\ge 1$)
\be\label{npn}
\frac{d}{dt}p(n,t) =\gamma(t,n-1) p({n-1},t) - \gamma(t,n) p(n,t).
\ee
This can be solved analytically via the moment generating function ${G}(\chi,t)$, \eq{Gunidef} which in the long--time limit reads  \cite{Bra10} 
\be
 G(\chi,t\to\infty) &=& e^{i\gamma t \chi +\frac{1}{g} \left( i\chi - \mbox{\rm Li}_2(1- e^{-i\chi}) \right)},
\ee
with the polylogarithm integral $\mbox{\rm Li}_2(z)\equiv \int_z^0 \frac{dt}{t} \ln (1-t)$.
Using the definition \eq{wtnscheme} for the waiting time distributions weighted with the probabilities $p_n(t)$ in the limit $t\to \infty$, we obtain
\be\label{wntfull}
w(\tau) &=& -\lim_{t\to \infty} \sum_{n=0}^\infty p_n(t) \frac{d}{d\tau} e^{ -\gamma \tau \left[ 1 + g \left( \gamma \left( t +\frac{\tau}{2} \right)   - (n-1) \right) \right] } \nonumber \\
&=& - \lim_{t\to \infty} \frac{d}{d\tau}     G(\chi = -i g\gamma \tau      ,t  )     e^{ -\gamma \tau \left[ 1 + g \left( \gamma \left( t +\frac{\tau}{2} \right)   +1) \right) \right] }         \nonumber   \\
&=&  -  \frac{d}{d\tau} e^{-f_g(\tau)}\nonumber\\
f_g(\tau) &\equiv&   g \gamma \tau \left[ \frac{\gamma \tau}{2}    +1 \right]    + \frac{1}{g}\mbox{\rm Li}_2(1- e^{- g\gamma\tau}).
\ee
For small $g\ll 1$, using Li$_2(z) \approx z$ we find  this to be very close to the Poissonian  waiting time distribution $w(\tau) \to \gamma e^{-\gamma \tau}$. Expansion up to first order in $g$ is valid for not too large $\tau$ and leads to \eq{wtFBexp}. Since $f_g(\tau)$ monotonously increases as a function of $\tau$, the  waiting time distribution \eq{wntfull} is always positive.

We note that formally, a Poissonian form also appears at very large $g\gg 1$ when scaling the time $\tau$ as  
$ \frac{1}{g}w(\tau/g) \to   \gamma e^{-\gamma  \tau }$, though the linear feedback model \eq{FBexperiment} for the rates becomes unrealistic then.

\section{Moment generating function for passive feedback model}
Here, we derive \eq{Gelectrons} for the passive control system in the model in  Fig. (\ref{waitingtime_fbscheme}). The Liouvillian $\mathcal{L}(\chi,\eta)$
in the space of states $0,...,N$ then reads
\be
\mathcal{L}(\chi,\eta)=\gamma_N\left( \begin{array}{ccccc}  
-1  &        0  & 0    &  ...    & e^{i\chi+i\eta} \\ 
 e^{i\eta}  &        -1 & 0    &  ...    & 0        \\ 
 0  & e^{i\eta} & -1   & 0    &  ...        \\ 
  ...  &  ...         & ...  & ...   &  ...        \\ 
0 &  ...&  0 &  e^{i\eta}  &   -1  \end{array} \right).
\ee
The moment generating function for counting to start at time $t=0$ in Laplace space obtained as
\be
\hat{G}(\chi,\eta,z) &=& \bra{0} \left(z-\mathcal{L}(\chi,\eta)\right)^{-1} \ket{0}\\
&=& \sum_{i=1}^{N+1}  \left(z-\mathcal{L}(\chi,\eta)\right)^{-1}_{i1},
\ee
where we used the (empty) state $0$ as initial condition. Here, the inverse can be calculated explicitly, using the determinant
\be\label{Gdeterminant}
\det  \left(z-\mathcal{L}(\chi,\eta)\right) = (z+\gamma_N)^{N+1} -e^{i\chi} (\gamma_Ne^{i\eta})^{N+1} 
\ee
and the formula for the inverse of a matrix with expressions for the adjuncts that can be easily derived. The result is given by 
\be\label{GChieta}
\hat{G}(\chi,\eta,z) &=& \frac{(z+\gamma_N)^{N+1} - (\gamma_Ne^{i\eta})^{N+1}    }{(z+\gamma_N)^{N+1} -e^{i\chi} (\gamma_Ne^{i\eta})^{N+1} }\nonumber\\
&\times& \frac{1}{z+\gamma_N(1-e^{i\eta})}. 
\ee
If we are interested in a reduced description, one of the counting field can be set to zero. When only counting phonons, we have the simple expression
$
\hat{G}(\chi=0,\eta,z) = \left({z + \gamma_N ( 1- e^{i\eta})}\right)^{-1}
$
corresponding to a Poissonian process of phonon emission at rate $\gamma_N$. 
When only counting electrons, we have to set $\eta=0$ and obtain \eq{Gelectrons} with \eq{waithard}.

The long--time behavior of ${G}(\chi,\eta,t)\propto e^{t z_0}$ in the time domain is determined by the zero $z_0(\chi,\eta)$ in \eq{Gdeterminant} as the denominator of the first factor in \eq{GChieta},
$z_0(\chi,\eta) =  \gamma_N \left( e^{i\eta}  e^{i  \chi/(N+1)} -1\right)$, from which \eq{Gchietalongtime} follows.

\end{appendix}

\end{document}